\documentclass[aps,prf,twocolumn,superscriptaddress,9pt,usenames,dvipsnames]{revtex4-1}
\usepackage{amsmath}
\usepackage{graphicx}
\usepackage{mathtools}
\usepackage{mathtools}
\usepackage{stmaryrd}
\usepackage{float}
\usepackage{empheq}
\usepackage{epstopdf}
\usepackage{bigints}
\usepackage{epstopdf}
\usepackage{amssymb}
\usepackage{wrapfig}
\usepackage{amsmath,hyperref}
\usepackage[noabbrev]{cleveref}
\usepackage{soul}
\usepackage{textcomp}
\usepackage{mydef}
\usepackage{xcolor}
 \usepackage{comment}
\usepackage[makeroom]{cancel}

\hypersetup{
colorlinks,
citecolor=blue,
filecolor=red,
linkcolor=blue,
urlcolor=blue,
hyperfigures
}

\begin{document}

\title{Growing, Buckling, and Swirling: motility from polymerization}

\author{Naveen Kumar D}
\affiliation{International Centre for Theoretical Sciences, Tata Institute of Fundamental Research,
Shivakote, Bengaluru 560089, India}

\author{Michael J. Shelley}
\affiliation{Center for Computational Biology, Flatiron Institute, 
	New York, NY 10010, USA; 
	Courant Institute of Mathematical Sciences, 
	New York University, New York, NY 10012, USA}

\author{Brato Chakrabarti}
\email{brato.chakrabarti@icts.res.in} 
\affiliation{International Centre for Theoretical Sciences, Tata Institute of Fundamental Research,
Shivakote, Bengaluru 560089, India}

\vspace*{-1cm}
\date{\today}

\begin{abstract}
Locomotion in low-Reynolds-number environments is achieved through a remarkable diversity of strategies, from flagellar rotation and ciliary beating to large-scale body deformations.  A distinct and biologically important class of propulsion arises when surface-anchored filaments grow and collectively reorient — as seen in the cellulose-extruding bacterium \textit{Acetobacter xylinum} and in recent experiments on actin-propelled synthetic colloids inspired by the motility of \textit{Listeria monocytogenes} — suggesting that polymerization itself is a generic route to self-propulsion. Developing a theoretical framework for this class of problems requires simultaneously resolving filament kinetics, their orientational dynamics, and fluid-structure interactions — all self-consistently coupled to the resulting locomotion. To address this, we formulate a continuum framework in which the active forces driving locomotion emerge self-consistently from filament nucleation, growth, catastrophe, and hydrodynamic interactions. We show analytically that polymerization-induced compressive forces drive a long-wavelength buckling instability, leading to spontaneous symmetry breaking of the filament carpet and large-scale flows. In coupling this framework to a force- and torque-free motile spheroidal particle, a wide variety of behaviors emerge -- this includes spontaneous spinning, directed motility, and chiral swimming -- whose selection is governed by the spatial patterning of polymerizing filaments. These results establish a general theoretical foundation for motility, driven by collective dynamics of polymerizing filaments and point towards new design principles for synthetic micron-scale swimmers.
\end{abstract}
\pacs{...}
\maketitle

\section{Introduction}

The world of swimming microorganisms is governed by physics strikingly different from our everyday experience \cite{10.1119/1.10903}. Microorganisms such as bacteria, algae, and protozoa navigate fluid environments where inertial forces are negligible compared to viscous forces — a regime described by the incompressible Stokes equations: 
\begin{equation}\label{eq:Stokes_equation}
	-\nabla q + \mu \Delta \mathbf{u} = \mathbf{0}, \quad \nabla \cdot \mathbf{u} = 0, 
\end{equation}
where $\mu$ is the viscosity, $\bu$ is the velocity field, and $q$ is the pressure. These are linear, constant-coefficient, elliptic PDEs with no explicit time dependence — the instantaneous flow is entirely determined by the instantaneous boundary conditions, and as a result, the governing equations are time-reversible \cite{Batchelor_2000}. Breaking this time-reversibility is a prerequisite for locomotion, and microorganisms have evolved a remarkable diversity of strategies to do so: the rotation of helical flagella in bacteria such as \textit{E. coli} \cite{berg2003rotary}, the  beating of flagella in sperm cells \cite{doi:10.1098/rspa.1951.0218}, coordinated metachronal waves of cilia in organisms such as \textit{Paramecium} \cite{doi:10.1073/pnas.1218869110,Machemer1972}, and large-scale body deformations in amoeboidal swimmers \cite{AOUN20201157,PhysRevLett.111.228102}. Understanding the physical principles underlying these strategies and the flows they generate is a central problem in biological fluid mechanics.

\begin{figure}
    \centering
    \includegraphics[width=1\linewidth]{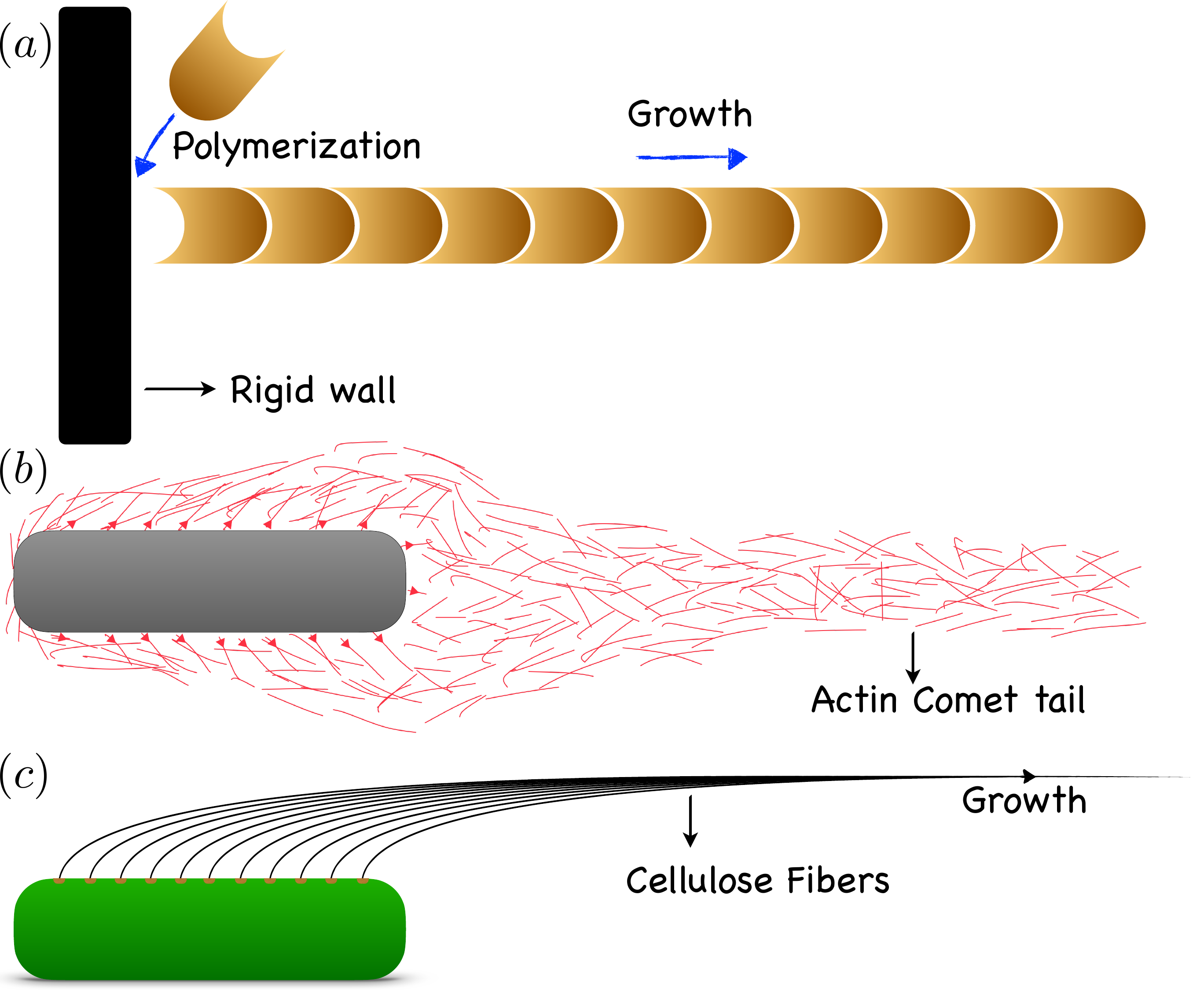}
    \caption{(a) Schematic illustrating force generation due to filament polymerization against a rigid surface. Monomers are added at the anchored end of the filament, driving polymerization that pushes the filament into the surrounding fluid. The viscous resistance of the medium can then lead to compressive forces along the filament backbone. (b) Schematic of actin polymerization on the surface of \textit{Listeria monocytogenes}. Polymerizing filaments form a tail-like structure, known as an actin comet tail, which propels the bacterium through the cytoplasm. (c) Schematic of cellulose fiber extrusion from the Gram-negative bacterium \textit{Acetobacter xylinum}. The extending fibers exert a reactive force on the cell body, contributing to its motility.}
    \label{fig:motivation}
\end{figure}

A particularly well-studied class of microswimmers achieves propulsion through surface-driven flows. Active processes localized near the organism's surface generate forces that stir the surrounding fluid, which in turn results in the net motion of the body. Given the boundary-value nature of the Stokes flow, the surface activity entirely determines the resulting locomotion — a key insight that motivates the squirmer model, introduced by Lighthill \cite{Lighthill1952} and later extended by Blake \cite{blake1971spherical}. In these models, the effect of surface activity is represented by a prescribed tangential slip velocity on the swimmer's surface. Though originally motivated by ciliary carpets, the squirmer framework has proven useful for studying the far-field hydrodynamics of microswimmers \cite{annurev:/content/journals/10.1146/annurev-fluid-121021-042929}, their mutual interactions \cite{Ishikawa_Pedley_Drescher_Goldstein_2020}, and collective behavior in suspensions \cite{Chamolly_Ishikawa_2026}. Yet, such models take the boundary conditions as given, leaving open the question of how biophysical processes near the cell surface self-organize to produce active forces in the first place. An interesting and relatively unexplored direction is to ask how the dynamics of an anchored active layer on the cell surface, such as the beating cilia \cite{doi:10.1073/pnas.2113539119}, motor-filament assemblies \cite{Monteith2016, PhysRevLett.126.028103}, or polymerizing cytoskeletal filaments \cite{LIN20101043} — collectively give rise to the macroscopic forces that drive flows and locomotion.

In this work, we address this question in the context of a distinct and biologically important class of propulsion: motility driven by the growth of surface-anchored cytoskeletal polymers or filaments. Growing biopolymers can generate substantial mechanical forces even in the complete absence of molecular machines \cite{Howard2001}. In such situations, a filament anchored to a surface polymerizes as monomers are added at the anchoring site, displacing the filament outward and generating a polymerization-driven force that arises from the coupling between the free energy of monomer addition and the mechanical work required to overcome the load \cite{phillips2012physical} (see Fig.~\ref{fig:motivation}-{$(a)$}). For actin filaments, such forces are typically on the order of a few piconewtons per filament; for microtubules, they can reach $\mathcal{O}(10)$ piconewtons — forces sufficient to deform membranes, reposition organelles, and drive cell motility \cite{kumar2025forcesscalecell}. A well-known example of polymerization-driven motility is the intracellular pathogen \textit{Listeria monocytogenes} (see Fig.~\ref{fig:motivation}-$(b)$). Upon invading a host cell, \textit{Listeria} nucleates a branched, cross-linked actin network at its rear surface, where monomers are added at the filament ends near the bacterium. As a result, the cell is propelled through the cytoplasm, whereby the drag forces on the cell body are balanced by reactive forces transmitted through the polymerizing actin network, which forms the characteristic actin comet tail. While the \textit{Listeria} system has been extensively studied and modeled \cite{joanny2009active}, it operates within the highly confined, gel-like cytoplasm of a host cell where filaments form a cross-linked network — complementary to the surface-driven, Stokesian setting we consider.

A more direct biological motivation for the present work is provided by the Gram-negative bacterium \textit{Acetobacter xylinum} (see Fig.~\ref{fig:motivation}-$(c)$). The bacterium synthesizes cellulose fibers that are extruded through pores distributed along its surface and polymerize directly into the surrounding fluid \cite{ffffa456249a4bed89485e12d11cd89c}. Monomers are added at the anchoring site on the bacterial surface, so that the fiber tip is displaced outward as the filament grows, with the resulting fibers exceeding the bacterium's own body size. As polymerizing cellulose fibers extend into the fluid, they eventually exceed a critical length at which the compressive viscous forces resisting filament growth are sufficient to cause buckling \cite{ffffa456249a4bed89485e12d11cd89c}. This elastic instability reorients the filament tip, redirecting the force it exerts on the surrounding fluid. The interplay between polymerization, buckling, and hydrodynamic interactions among the many fibers distributed over the cell surface gives rise to collective dynamics whose consequences for motility remain poorly understood \cite{Cannon2000}.

More broadly, the question of how a dense carpet of surface-anchored, polymerizing filaments interacts through mechanical forces and fluid flows and possibly drives locomotion via self-organized dynamics is an interesting modeling problem at the interface of fluid dynamics and non-equilibrium physics. The challenge is substantial: unlike ciliary carpets, where the beating pattern can be prescribed, here the filament orientations and hence the active forces must be determined self-consistently with the flow they generate. Such a framework requires simultaneously resolving the kinetics of polymerization, the orientational dynamics of a dense filament bed, and the fluid-structure interaction with the surrounding Stokesian fluid. The importance of such a framework is underscored by recent experiments demonstrating that synthetic colloids coated with actin nucleation-promoting factors can be propelled by actin treadmilling, with directed self-propulsion and collective flocking emerging spontaneously from the filament dynamics alone \cite{Lopes2025EmergenceOA} — suggesting that polymerization-driven motility is not merely a biological curiosity but can also be a viable design principle for synthetic microswimmers \cite{mi11121048, GANGULY2026204529}

Taken together, these observations motivate the central question we address here: given a surface patterned with polymerizing cytoskeletal filaments, what instabilities, self-organized flows, and motility can emerge purely from the growth and reorientation of anchored filaments? The remainder of the paper is organized as follows. In Sec. II, we develop the mean-field continuum model for a polymerizing filament carpet, deriving the polarization dynamics and the coarse-grained active forces. A linear stability analysis in a half-space geometry is carried out in Sec. III revealing a long-wavelength buckling instability. The computational framework is described in Sec. IV. Nonlinear simulation results of the full problem are presented in Sec. V. We conclude with a discussion of the results and outlook in Sec. VI.

\section{Mean field model for polymerizing active carpet}\label{sec:model}
Motivated by several examples discussed in the introduction, here we consider the dynamics of a dense carpet of polymerizing filaments. Owing to the large number of filaments involved, we adopt a mean-field description and introduce a set of simplifying assumptions to capture the essential features of the dynamics. In this section, we outline the key ingredients of this mean-field model.

\subsection{Nucleation and growth of polymers} \label{subsec:FP_without_p}

We begin by focusing on the kinetics of polymerization of biopolymers, of which microtubules (MTs) \cite{Howard2001} are a prominent example. In the present context, we consider that these filaments undergo three key processes: (i) nucleation on a surface, (ii) continuous growth via polymerization at a constant velocity $V_g$, as subunits are added from an effectively infinite monomer bath, and (iii) depolymerization through catastrophe events. Cytoskeletal filaments are intrinsically polar and, as outlined in the introduction, we assume that monomers are added at the nucleation site on the surface, so that growth requires displacement of the polymerizing filament.  
\begin{figure}[H]
    \centering
    \includegraphics[width=1\linewidth]{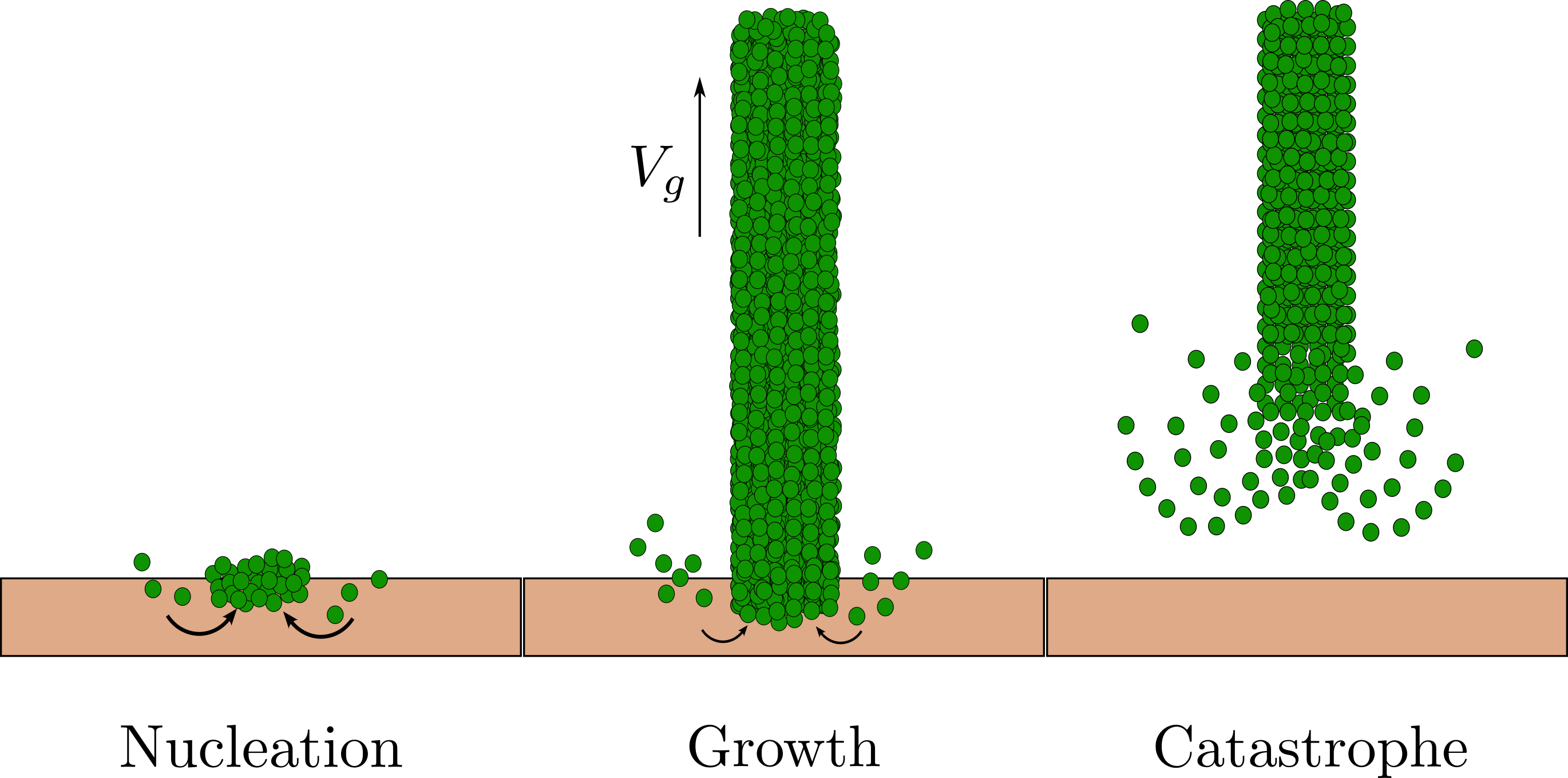}
    \caption{Schematic of filament polymerization kinetics. (Left) Spontaneous nucleation on the surface, driven by local monomer accumulation. (Middle) Steady growth through polymerization at a constant velocity $V_g$, with monomers added at the surface-anchored end. (Right) Catastrophe event illustrating stochastic depolymerization and filament disassembly.}
    \label{fig:polymerization}
\end{figure}
In a continuum description, this kinetics can be modeled by a distribution function $\Psi(\ell,t)$ that serves as a measure of probability density of finding a filament of length $\ell$ at time $t$ and is normalized as $\int \Psi(\ell,t) \md \ell = N(t)$, where $N(t)$ is the total number of filaments \cite{10.7554/eLife.55877}. The evolution of $\Psi(\ell,t)$ obeys a conservation equation
\begin{equation}\label{eq:fp0}
    \partial_t \Psi + \partial_\ell (V_g \Psi) = -\lambda \Psi.
\end{equation}
Although nucleation and catastrophe are microscopically Poisson processes, the continuum description retains only their mean-field consequences: growth enters deterministically as a drift in length space, while catastrophe enters as the
sink set by the mean rate $\lambda$. Nucleation likewise appears only through its mean rate $\gamma$, not in the bulk equation but as a boundary condition, as we now illustrate. Integrating the above equation over filament length yields an evolution equation for the total number of filaments in the carpet,
\begin{equation}\label{eq:logistic equation}
    \frac{\md N}{\md t} = \gamma - \lambda N.
\end{equation}
This balance follows from the nucleation condition $V_g \Psi(\ell_0,t) = \gamma$, which sets the flux of new filaments into the population at the minimum length $\ell_0$. At steady state, one then obtains $N_\text{steady} = \gamma/\lambda$ and an exponential steady-state length distribution,
\begin{equation}\label{eq:psist}
	\Psi_{\mathrm{steady}}(\ell)
	= \frac{\gamma}{V_g}
	\exp\!\left[-\frac{\lambda}{V_g}(\ell-\ell_0)\right],
\end{equation}
with mean filament length
$\bar{\ell} = V_g/\lambda + \ell_0$. In what follows, averaging over length distributions will be denoted as:
\begin{equation}
\overline{(\cdot)}=\int_{\ell_0}^\infty(\cdot)\Psi_{\mathrm{steady}}(\ell)\mathrm{d}\ell.
\end{equation}

\subsection{Fluid flows and isolated filament dynamics}
\begin{figure*}[hbtp]
    \centering
    \includegraphics[width=1\linewidth]{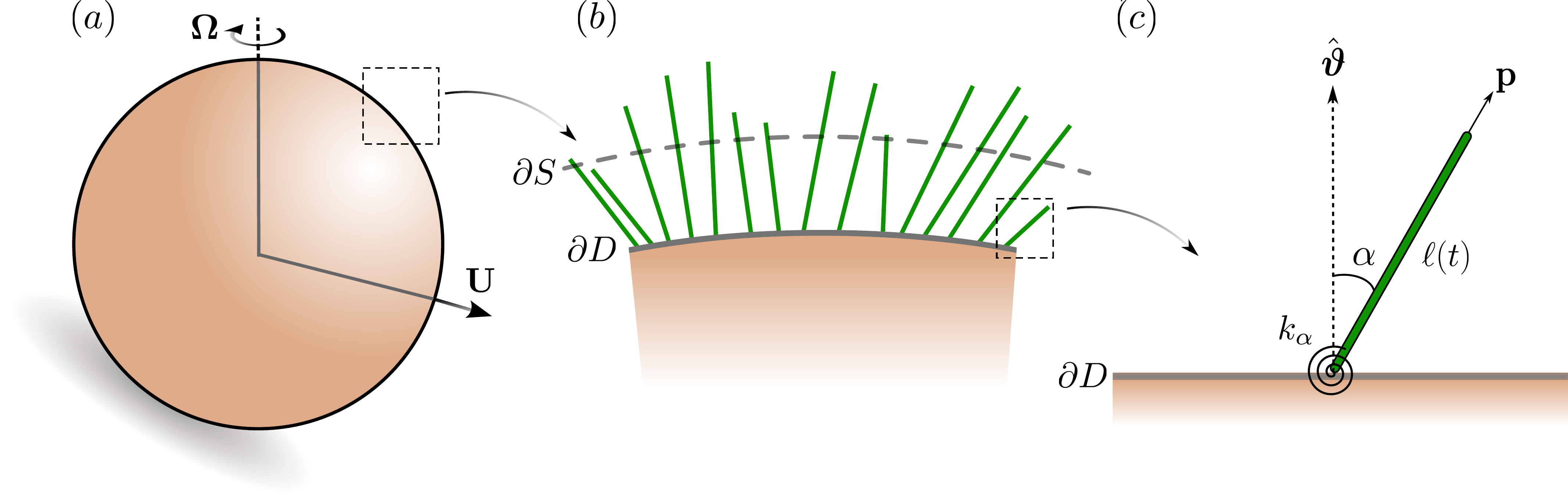}
    \caption{Model of a mobile rigid body moved by a growing active carpet.
    (a) The body translates with velocity $\mathbf{U}$ and rotates with angular velocity $\boldsymbol{\Omega}$.
    (b) A carpet of filaments with varying lengths emerges from the surface $\partial D$; the dashed surface $\partial S$ represents the envelope of their average length.
    (c) Close-up of a single filament modeled as a rigid rod of length $\ell(t)$ and orientation $\mathbf{p}$, anchored to the surface by a torsional spring of stiffness $k_\alpha$, which penalizes deviations from the local surface normal $\hat{\boldsymbol{\vartheta}}$.} 
    \label{fig:rod_approximation}
\end{figure*}

We now turn to ask how fluid flows generated by an active carpet couple to filament dynamics. To this end, we consider a carpet on the surface of a rigid body submerged in a Stokesian fluid (see Fig.~\ref{fig:rod_approximation}-$(a)$,$(b)$) governed by the incompressible Stokes equations given in Eq.~\eqref{eq:Stokes_equation}. The carpet is anchored to the surface of a rigid body $\partial D$, parameterized by a surface coordinate $\by$. The Stokes equations are supplemented by the no-slip condition on $\partial D$,
\begin{align}
	\bu\big|_{\partial D} = \bv(\by) = \bU + \boldsymbol{\Omega} \times (\by-\bx_0),
\end{align}
where $\bv(\by)$ denotes the velocity of the surface at the point $\by$, with $\bU$ and $\boldsymbol{\Omega}$ being the translational and angular velocities of the rigid body, respectively, and $\bx_0$ is the geometric center of the body. If the rigid body is stationary, one has $\bv(\by) = \mathbf{0}$; in the circumstance where the body is free to move, we will have to determine $\{\bU, \boldsymbol{\Omega}\}$ self consistently as a part of the solution in conjunction with the condition that the object is force and torque free. The Stokesian flow is driven by forces generated from filament polymerization on the surface. The precise manner in which a coarse-grained driving emerges from a polymerizing carpet will be derived in Sec.~\ref{sec:coarse_graining}. For now, we note that the resulting fluid motion modifies filament orientations, which in turn reshapes the forces and driving forces responsible for the flow within the active carpet. To formalize this micro--macro coupling, we begin by considering the dynamics of an isolated filament. 

To keep the problem tractable within a mean-field framework, we model each filament as a rigid rod of time-dependent length $\ell(t)$ and orientation $\mathbf{p}$. The rod grows uniformly along its axis at a constant speed $V_g$, starting from a fixed length $\ell_0$ at nucleation. We assume that filaments are nucleated along the local surface normal $\hat{\boldsymbol{\vartheta}}$ of the rigid body surface $\partial D$. To mimic the bending elasticity of anchored filaments and possible compliance at the anchoring site, each rod is equipped with a torsional spring of stiffness $k_\alpha$, which resists deviations of the filament orientation from the local surface normal \cite{doi:10.1073/pnas.2405114121} (see Fig.~\ref{fig:rod_approximation}-$(c)$). This spring  generates a torque $k_\alpha \mathbf{T}_0$ about the anchoring point, where
\begin{equation}
    \mathbf{T}_0 = \cos^{-1}(\mathbf{p} \cdot \hat{\boldsymbol{\vartheta}})\,
    \frac{\mathbf{p} \times \hat{\boldsymbol{\vartheta}}}{\|\mathbf{p} \times \hat{\boldsymbol{\vartheta}}\|}.
\end{equation}
On using the slender-body theory (SBT) \cite{Batchelor_1970}, we can compute the force per unit length exerted by a filament on the fluid as
\begin{equation}\label{eq:sbt}
	\mathbf{f}(\bx) = \frac{4\pi \mu}{\ln(2/\varepsilon)} \left( \mathbf{I} - \frac{\mathbf{p}\mathbf{p}}{2} \right) \cdot (\dot{\mathbf{x}}(s) - \mathbf{u}(\bx)),
\end{equation}
where $\dot{\mathbf{x}}$ is the velocity of a material point on the filament centerline parameterized by arc-length $s \in [0, \ell(t)]$;  $\bu(\bx)$  is a background mean-field flow, yet to be determined, and $\varepsilon = a/\ell \ll 1$ is the filament slenderness ratio. The center-line velocity can be further expressed as
\begin{equation}\label{eq:fil-vel}
	\dot{\mathbf{x}}(s,t) = \mathbf{v}(\mathbf{y}) + V_g \mathbf{p} + s\, \dot{\mathbf{p}}. 
\end{equation}
Here, the first term arises from the rigid-body motion of the anchoring point at $s=0$, the second term stems from polymerization, and the last term accounts for re-orientation dynamics around the anchoring point. Assuming that the emergent fluid velocity varies at length scales larger than $\ell(t)$, we can use a first-order approximation \cite{doi:10.1073/pnas.2405114121} to write
\begin{equation}\label{eq:fluid-vel}
	\mathbf{u}(\bx(s)) = \mathbf{v}(\mathbf{y}) + s\, \mathbf{p} \cdot \nabla \mathbf{u}\big|_{s=0}, 
\end{equation}
where the no-slip condition ensures $\mathbf{u}(\bx(0)) = \mathbf{v}(\mathbf{y})$. Upon using the above expression in Eq.~\ref{eq:sbt} and exploiting torque balance around the anchoring point at $s=0$, we obtain
\begin{align}
	\dot{\mathbf{p}} &= \left( \mathbf{I} - \mathbf{p}\mathbf{p} \right)
	\cdot \nabla \mathbf{u}\Big|_{s=0} \cdot \mathbf{p}
	+ \frac{k_\alpha}{\xi_r(\ell)}\, \mathbf{T}_0 \times \mathbf{p},  \label{eq:Jeffery} \\
	    \mathbf{f}(s, \mathbf{p}) &= \frac{4\pi \mu s}{\ln(2/\varepsilon)} \left[ \frac{k_\alpha}{\xi_r(\ell)} \left( \mathbf{T}_0 \times \mathbf{p} \right) - \frac{\mathbf{p}\mathbf{p}\mathbf{p}}{2} : \nabla \mathbf{u}\Big|_{s=0} \right]\nonumber \\
        &+ \frac{2\pi\mu V_g \mathbf{p}}{\ln(2/\varepsilon)},\label{eq:force-s-p}
\end{align}
where $\xi_r(\ell) = \dfrac{4\pi \mu \ell^3}{3 \ln(2/\varepsilon)}$ is a rotational drag coefficient. In Eq.~\eqref{eq:Jeffery}, the first term is the so-called Jeffery’s equation for a rigid rod in Stokes flow~\cite{jeffery1922motion} that describes reorientation of a rod-like object in locally linear flow fields,  and the second term represents reorientation due to the restoring torque. In Eq.~\eqref{eq:force-s-p}, the first term highlights the response of the anchored rod in a flow-field and the associated force-density results from the rigidity constraint of the centerline; here the tensor contraction $\mathbf{p}\mathbf{p}\mathbf{p}: \nabla \mathbf{u}$ denotes $p_i p_j p_k \partial_k u_j$. The second term of Eq.~\eqref{eq:force-s-p} is an active force resulting from polymerization of the filament and bears resemblance to polar force densities exerted by cargo-carrying molecular machines along the filament backbone in anchored cytoskeletal fiber beds \cite{Dutta2024, doi:10.1073/pnas.2405114121}.

\subsection{Collective dynamics in the carpet}\label{sec:coarse_graining}

In our description of the polymerization kinetics in \ref{subsec:FP_without_p}, we have deliberately neglected orientational dynamics and spatial heterogeneity of the growing filaments for simplicity. Building on our understanding of the orientational dynamics of an isolated filament in the carpet, we now extend the mean-field framework to account for filament reorientation in a heterogeneous active carpet. Here, we describe the carpet by a generalized distribution function  $\psi(\mathbf{y}, \ell, \mathbf{p}, t)$ that now measures the probability density of finding a filament of length $\ell$, with orientation $\mathbf{p}$ at a surface coordinate $\mathbf{y}$ at time $t$. Evolution of $\psi$ is governed by a Fokker-Planck equation
\begin{equation}\label{eq:fokker-planck}
    \partial_t \psi + V_g \, \partial_\ell \psi + \partial_{\mathbf{p}} (\dot{\mathbf{p}} \, \psi) = -\lambda \psi, 
\end{equation}
that differs from Eq.~\eqref{eq:fp0} due to the presence of rotational velocity, $\dot{\bp}$ given by Eq.~\eqref{eq:Jeffery}; here $\partial_\bp = (\mathbf{I} - \bp \bp) \cdot \nabla_\bp$ denotes the surface gradient operator on the unit sphere. Equation~\eqref{eq:fokker-planck} is accompanied by the following boundary conditions
\begin{align}
    \psi(\mathbf{y},\ell \rightarrow\infty,\mathbf{p},t)&=0, \label{eq:BC-1}\\
    \psi(\mathbf{y},\ell_0,\mathbf{p},t)&=c(\mathbf{y})\dfrac{\gamma}{V_g}\,\delta\!\left[\mathbf{p}-\hat{\boldsymbol{\vartheta}}(\mathbf{y})\right], \label{eq:BC-3}
\end{align}
where $c(\mathbf{y}) = \bar{\rho}(\by) c_0$, is the local surface density of nucleation sites, where $c_0 = n_c/ A_s$, with $A_s$ being the area of the anchoring surface and $n_c$ being the total number of nucleation sites. Here, $\bar{\rho}(\by)$ accounts for possible density fluctuations or heterogeneity on the surface, normalized such that $\int_{\partial D} \bar{\rho}(\by) = 1$. Physically, such heterogeneity may arise from pre-patterned nucleation sites on the surface of the body. In summary, Eq.~\eqref{eq:BC-1} accounts for the finite length of filaments and Eq.~\eqref{eq:BC-3} encodes the nucleation condition, ensuring consistency with the evolution of the net number of filaments and imposing that filaments are nucleated normal to the substrate.

It is convenient to define now an orientation-marginalized distribution function $\Psi(\by,\ell,t)$ that obeys Eq.~\eqref{eq:fp0}. Upon using Eq.~\eqref{eq:BC-3}, it is easy to see that in the steady-state, the marginalized distribution function is given by
\begin{equation}
\begin{split}
    \Psi_\text{sm}(\by,\ell) &= c(\mathbf{y}) \frac{\gamma}{V_g} \exp\left[-\frac{\lambda}{V_g}(\ell - \ell_0)\right] \\
    &\equiv c(\by) \Psi_\text{steady}(\ell).
\end{split}
\end{equation}
The above expression is identical to $\Psi_\text{steady}(\ell)$ obtained in Eq.~\eqref{eq:psist}, now modulated by the spatial heterogeneity in polymerization through the surface density $c(\mathbf{y})$.  In this work, we focus on a regime in which the kinetics of nucleation and polymer growth occur on time scales that are fast compared to the reorientation dynamics of the filaments. This separation of time scales is biologically likely: the characteristic reorientation time scale of a $\bar{\ell} \sim 20 \ \mu$m long MT is $\tau_r \sim \xi_r/k_\alpha \sim \mathcal{O}(10^3)\,\mathrm{s}$ which is well separated from the polymerization scale,
$\lambda^{-1} \sim \mathcal{O}(20$-$30)\,\mathrm{s}$. Motivated by this scale separation, we adopt the Ansatz
\begin{equation}\label{eq:appfp}
    \psi \approx  \Psi_\text{sm}(\by,\ell) \tilde{g}(\mathbf{y}, \mathbf{p},t),
\end{equation}
where the function $\tilde{g}$ is effectively independent of $\ell$, up to weak variations required to satisfy the boundary conditions. Physically, the rapid equilibration of filament-length distributions allows adiabatic elimination of polymer growth and decay, leaving the slow orientational dynamics governed primarily by fluid flow and bending elasticity.

Next, we highlight that evolving the high-dimensional distribution function $\psi(\mathbf{y},\ell,\mathbf{p},t)$ is often costly, and it is convenient to consider the dynamics of its orientational moments, which admit a more direct physical interpretation. To this end, we define a polarization field $\mathbf{n}(\mathbf{y},t)$, representing the local mean orientation or magnetization of filaments anchored at the surface point $\mathbf{y}$, as
\begin{equation}\label{eq:polarity}
    \mathbf{n}(\mathbf{y},t) = \frac{1}{c(\by)} \int_\mathcal{H} \int_{\ell_0}^\infty \mathbf{p}\,\psi\,\mathrm{d}\ell\,\mathrm{d}\mathbf{p},
\end{equation}
where $\mathcal{H}$ is the unit half-sphere. Upon using Eq.~\eqref{eq:force-s-p}, we obtain the evolution for the polarization field as
\begin{equation}
	\partial_t\mathbf{n}(\mathbf{y},t)
	= -\lambda \mathbf{n} + \lambda \hat{\boldsymbol{\vartheta}}(\mathbf{y})
	+ \frac{1}{c(\mathbf{y})}
	\int_\mathcal{H} \int_{\ell_0}^\infty \dot{\mathbf{p}} \, \psi \, \mathrm{d}\ell \, \mathrm{d}\mathbf{p}.
\end{equation}
The first term represents the loss of orientational coherence due to catastrophe or depolymerization events, which reset filament orientation and act as an effective sink. This contribution is analogous to rotational diffusion in suspensions of Brownian rods, where randomizing processes relax orientational order. The second term captures the spontaneous generation of polarization along the local surface normal through nucleation events. The final term encodes the deterministic reorientation of filaments driven by fluid flows and elastic torques. On using Eq.~\eqref{eq:Jeffery}, we can simplify the last term to obtain
\begin{equation}
	\partial_t \bn  =\lambda(\hat{\boldsymbol{\vartheta}}- \mathbf{n}) + \left[ \mathbf{I}\mathbf{n} - \langle \mathbf{ppp} \rangle\right] : \nabla \mathbf{u}\Big|_{\partial D} + \frac{1}{\tau_R} \Big\langle \mathbf{T}_0 \times \bp \Big \rangle.
\end{equation}
Here,  $\tau_R$ is an effective relaxation time for the carpet, defined by averaging the torsional relaxation rate over the steady-state length distribution,
\begin{equation}
	\tau_R^{-1} = \overline{\left( \frac{k_\alpha}{\xi_r(\ell)} \right )}.
\end{equation}
We have also introduced the following notation for orientational average
\begin{equation}
	\langle (.) \rangle(\by,t) = \int_\mathcal{H} (.)(\by,\bp,t) \tilde{g}(\by,\bp,t) \md \bp,
\end{equation}
where $\tilde g$, as defined earlier, is the (effective) orientational distribution obtained after adiabatic elimination of the filament-length dynamics. To close the micro--macro framework, we now have to compute the mean-field fluid velocity $\mathbf{u}(\mathbf{x})$ generated by the polymerizing carpet. Following Refs.~\cite{doi:10.1073/pnas.2405114121,PRXLife.3.023007}, we coarse-grain the active forcing as,
\begin{equation}
    \mathbf{F}_a(\by,t) = \int_{\mathcal{H}} \int_{\ell_0}^{\infty} \psi \ \left(\int_0^\ell \mathbf{f}(\by,s,\mathbf{p}) \md s\right )\md \ell \ \md \bp,
\end{equation}
where $\bff(\by,s,\bp)$ is the force density given in Eq.~\eqref{eq:force-s-p}. We model this coarse-grained active forcing as an effective traction jump acting across a fixed interface $\partial S \equiv \partial D + \bar{\ell}\,\hat{\boldsymbol{\vartheta}}$, obtained by offsetting the rigid surface $\partial D$ along its outward
normal (see Fig.~\ref{fig:rod_approximation}-$(b)$); here $\bar{\ell} = \ell_0 + V_g/\lambda$, is the mean filament length in the carpet, as introduced earlier. This approximation replaces the distributed volumetric forcing produced by the carpet with an equivalent surface traction applied at a representative height above the substrate. This approximation is suitable when the characteristic size of the body is much larger than the average filament length: a limit that we focus on in this problem. Such a traction-layer representation originates in early continuum models of ciliary carpets \cite{keller1975traction} and has since been successfully adapted to describe ciliary synchronization \cite{ishikawa2020stability, kanale2022spontaneous} and cytoplasmic flows \cite{doi:10.1073/pnas.2113539119}, where it has been shown to provide quantitative agreement with experiments \cite{PRXLife.3.023007}. In the present context, this approach provides a natural and tractable computational model that connects the microscopic filament forces to the macroscopically generated flow field.

\subsection{Non-dimensionalization and governing equations}
We non-dimensionalize the governing equations using the effective relaxation time $\tau_R$ as the time scale, the average filament length $\bar{\ell}$ as the length scale, and a viscous force scale $\mu/\tau_R$ for forces. The dimensionless polarization dynamics is then given as
\begin{equation}\label{eq:nev}
	\partial_t \bn  = \bar{\lambda} (\hat{\boldsymbol{\vartheta}}- \mathbf{n}) + \left[ \mathbf{I}\mathbf{n} - \langle \mathbf{ppp} \rangle\right] : \nabla \mathbf{u}\Big|_{\partial D} + \left\langle \mathbf{T}_0 \times \bp \right \rangle,
\end{equation}
where $\bar{\lambda} = \lambda \tau_R$ is a dimensionless catastrophe rate; we have used the same symbols as before to indicate dimensionless variables for notational simplicity.  In the same dimensionless units, the coarse-grained traction jump takes the form
\begin{equation}\label{eq:factive}
	 \mathbf{F}_a = \rho(\by) \bar{\rho} \left[\bar{\sigma} \bn  +  \chi \langle \mathbf{T}_0 \times \bp \rangle - \frac{\beta}{2} \langle \bp \bp \bp \rangle : \nabla \bu\Big|_{\partial D} \right],
\end{equation}
where $\rho(\mathbf{y})$ is the normalized density heterogeneity field on the surface and $\bar{\rho}$ sets the overall dimensionless surface density. We have introduced the following dimensionless groups:
\begin{itemize}
	\item Filament aspect ratio: $\eta = 2 \pi/\log(2/\varepsilon)$.
	\item Filament density: $\bar{\rho} = c_0 \left(\bar{\ell}\right)^2$.
	\item Activity from polymerization: \\ \vspace*{2mm} $\bar{\sigma} =  \frac{\eta V_g \tau_R}{\bar{\ell}} \sim \frac{\text{growth speed}}{\text{relaxational velocity}}$.
	\item Geometric factors: \\ \vspace*{2mm} $\chi  = \frac{3}{2} k_\alpha  \tau_R\overline{\left(1/\ell\right)}/{\mu  \left(\bar{\ell}\right)^2}$, \ \ $\beta = \eta\overline{\left(\ell^2\right)}/\left(\bar{\ell}\right)^2 $.
\end{itemize}
Along with $\bar{\lambda}$, the key control parameters in this model are the surface density $\bar{\rho}$ and the activity $\bar{\sigma}$. We note that the geometric factors $\chi$ and $\beta$ encode information about the filament length distribution; in a carpet with uniform filament lengths, both reduce to constants that depend on the filament slenderness.  We can now write the associated dimensionless Stokes problem as
\begin{alignat}{3}
	& -\nabla q + \Delta \mathbf{u} = \mathbf{0},
	\qquad
	 \nabla \cdot \mathbf{u} = 0,
	\\
	& \mathbf{u}(\mathbf{y})
	= \mathbf{U}
	+ \boldsymbol{\Omega} \times (\mathbf{y}-\mathbf{x}_0),
	\qquad
	&& \text{on } \partial D,
	\\
	& \llbracket \mathbf{u} \rrbracket = \mathbf{0},
	\quad
	\llbracket \boldsymbol{\sigma}\cdot\hat{\boldsymbol{\vartheta}} \rrbracket(\mathbf{y})
	= - \mathbf{F}_a(\mathbf{y}),
	\qquad
	&& \text{on } \partial S
	\equiv \partial D + \,\hat{\boldsymbol{\vartheta}}.
\end{alignat}
Here, $\boldsymbol{\sigma}\cdot\hat{\boldsymbol{\vartheta}}$ denotes the hydrodynamic traction, while ${\mathbf{F}}_{a}$ represents the force exerted on the rigid body by the anchored filament ends. In this formulation, the rigid body is allowed to move freely in the fluid; the special case of a stationary object is recovered by imposing $\mathbf{u}(\mathbf{y}\in\partial D)=\mathbf{0}$. To close the Stokes problem, we determine the unknown rigid-body velocities $\{\mathbf{U},\boldsymbol{\Omega}\}$ by enforcing global force- and torque-free conditions on the body:
\begin{align}
	\int_{\partial D} \left[\boldsymbol{\sigma} \cdot \hat{{\boldsymbol{\vartheta}}}+ 
	 \mathbf{F}_a  \right]  \md A &= \mathbf{0},\\
	\int_{\partial D} (\by-\bx_0) \times \left[\boldsymbol{\sigma} \cdot \hat{{\boldsymbol{\vartheta}}} + 
	 \mathbf{F}_a \right]   \mathrm{~d}A &=- \int_{\partial D} \langle \mathbf{T}_0 \rangle  \mathrm{~d}A  \label{eq:torqfree}.
\end{align}
The torque-free condition accounts for both the moment of these forces about the body center and the net torque generated by the torsional springs at the filament anchoring points. 

Taken together, Eqs.~\eqref{eq:nev}--\eqref{eq:torqfree} specify the complete micro--macro framework of our problem. Given a polarity field $\mathbf{n}(\mathbf{y},t)$, one can compute the active driving force via Eq.~\eqref{eq:factive}, which in turn determines the solution of the Stokes problem subject to the appropriate boundary conditions. The resulting velocity field then feeds back to reorient the filaments, thereby closing a self-consistent feedback loop. This approach presents several computational challenges: solving the Stokes equations around complex geometries, approximating higher-order orientational moments in terms of $\mathbf{n}$, and handling constraint forces in Eq.~\eqref{eq:factive} that depend implicitly on the unknown velocity field $\mathbf{u}$. We defer a discussion of the numerical methods used to address these challenges to Sec.~\ref{sec:methods}. We next analyze the model in a simple configuration to illustrate the possible emergent dynamics.

\section{Dynamics in half-space: linear stability analysis}\label{sec:linear_stability}

To gain physical insight into the onset of collective behavior and motion, we consider a half-space geometry in which the filament bed is anchored to a rigid planar surface at $z = 0$, bounding a semi-infinite Stokesian fluid domain occupying $z>0$. The system admits a trivial base state with a uniformly vertical filament orientation, $\mathbf{n}_0 = \hat{\mathbf{z}}$, and $\bu = \mathbf{0}$ corresponding to filaments that nucleate and grow normally from the surface.  In this state, no mean flow is generated, since the hydrostatic pressure in the fluid exactly balances the polymerization forces exerted by the filaments.

To analyze the stability of this state, we introduce a small orientational perturbation confined to the $x$–$z$ plane as $\mathbf{n}(\mathbf{x},t) \approx \hat{\mathbf{z}} + \delta \phi(t) \hat{\mathbf{x}}$, where $|\delta \phi| \ll 1$ represents the local tilt angle of the filaments. Linearizing the governing dynamics (Eqs.~\ref{eq:nev} and \ref{eq:factive}) for a spatially homogeneous perturbation yields the evolution equation
\begin{equation}\label{eq:homogeneous perturbation}
\partial_t \delta \phi = \left[ \bar{\rho}\bar{\sigma} - \left( \bar{\rho}\chi + 1 \right) - \bar{\lambda} \right] \delta \phi .
\end{equation}
In Eq.~\eqref{eq:homogeneous perturbation}, the first term, $\bar{\rho}\bar{\sigma}$, is destabilizing and originates from compressive forces generated by filament growth and their collective hydrodynamic interactions, which promote buckling. The second term, $\left( \bar{\rho}\chi + 1 \right)$, is stabilizing and arises from the elastic restoring torque of the filaments, which resists deviations from vertical alignment. The third term, $\bar{\lambda}$, is also stabilizing and stems from filament depolymerization; faster turnover of filaments prevents the persistence of bent configurations, and the associated hydrodynamics cannot lead to an orientational instability of the entire bed. The base state thus becomes unstable when
\begin{equation}
\bar{\sigma} > \chi + \frac{\bar{\lambda} + 1}{\bar{\rho}} .
\end{equation}
\begin{figure}[H]
    \centering
    \includegraphics[width=1\linewidth]{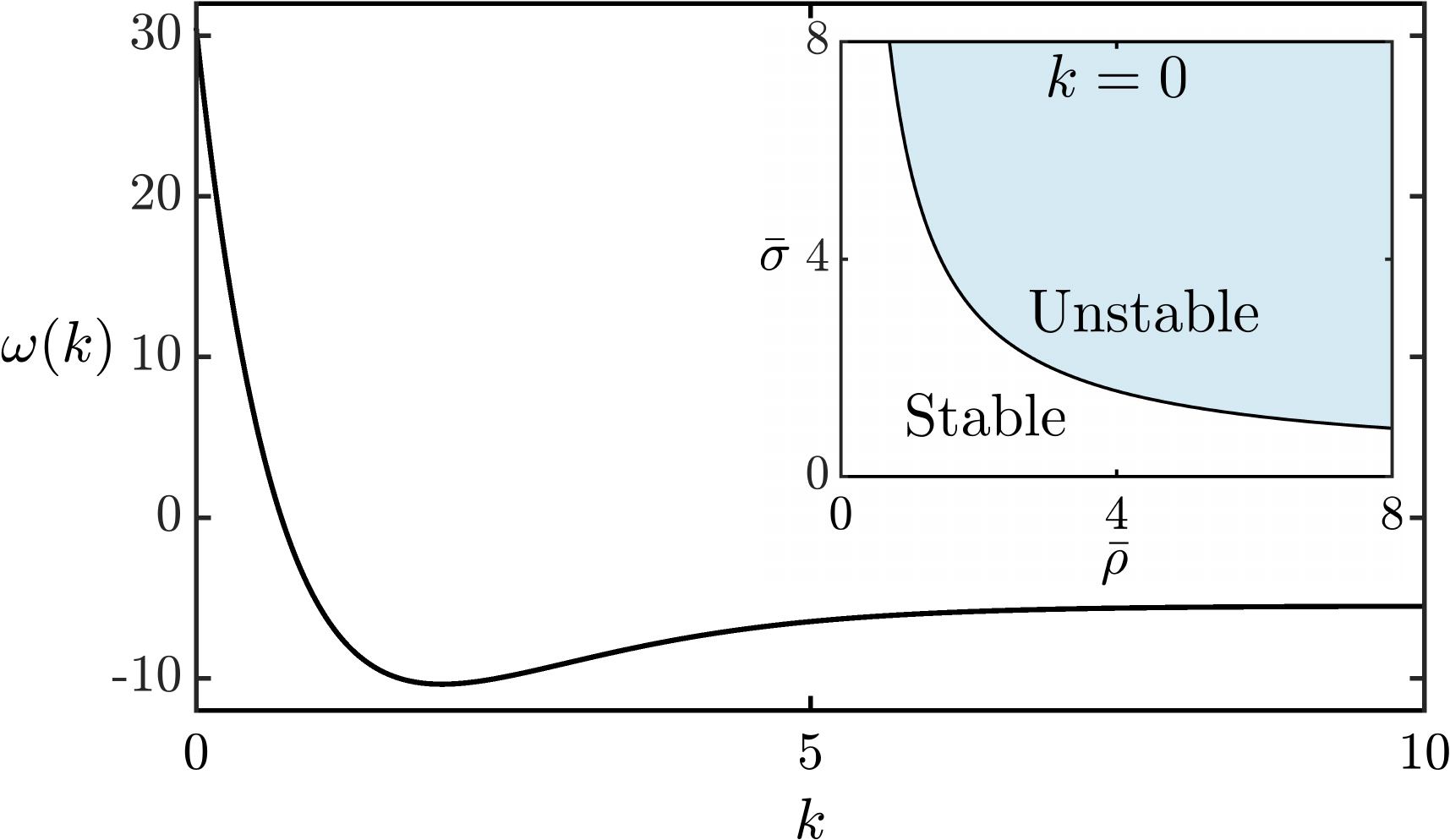}
    \caption{Linear stability analysis in a half-space about the base state $\mathbf{n}=\hat{\mathbf{z}}$.
    Dispersion relation showing the growth rate $\omega(k)$ as a function of wavenumber $k$ for representative parameters $(\bar{\lambda}=4.5,\, \bar{\sigma}=3.8,\, \bar{\rho}=10,\, \chi=0.2)$. The growth rate attains its maximum at $k=0$, indicating a long-wavelength (global buckling) instability.
    \textit{Inset:} Stability boundary at $k=0$ in the $(\bar{\rho},\bar{\sigma})$ parameter space. The shaded region denotes instability, showing that the base state loses stability with increasing $\bar{\rho}$ and $\bar{\sigma}$.}
    \label{fig:lsa}
\end{figure}
We next consider the response to spatial variations by analyzing a plane-wave perturbation of the form $\delta \phi(x,t) = \tilde{\phi}_k e^{ikx + \omega t}$; on solving the linearized problem, we find
\begin{equation}
\omega(k) = -(\bar{\lambda}+1) + e^{-k}(1-k)\bar{\rho}\left(-\chi+\bar{\sigma}\right).
\end{equation}
This dispersion relation reveals that the instability is long-wavelength, with $\omega(k)$ attaining its maximum at $k = 0$ whereby the entire polymerizing bed buckles uniformly. Furthermore, higher wavenumbers are exponentially damped (see Fig.~\ref{fig:lsa}), suggesting behavior that is different from the nematic elasticity of rods \cite{doi:10.1073/pnas.2405114121}, which typically shows algebraic decay.

\section{Computational methods}\label{sec:methods}

Guided by the linear stability analysis, we now proceed to numerical simulations of the full nonlinear equations. To make the problem computationally tractable, we employ two key approaches, which we outline in this section.

\subsection{Closure approximation}

The evolution of the polarity field $\mathbf{n}$ depends on the higher-order orientational moments such as $\langle \mathbf{pp} \rangle$, $\langle \mathbf{ppp} \rangle$, etc., and therefore requires a closure approximation to obtain a tractable system. To this end, we adopt a polar \textit{Bingham closure} following \cite{WEADY2022110937}, where the orientational distribution is assumed to take the form
\begin{equation}\label{eq:vMF dist}
\tilde{g}(\mathbf{p}) = \dfrac{1}{Z} \exp(\mathbf{b} \cdot \mathbf{p}),
\end{equation}
where $\mathbf{b}$ is a vector parameter that characterizes the degree of orientational bias, and 
\[
Z = \int_{\mathcal{H}} \exp(\mathbf{b} \cdot \mathbf{p}) \, \mathrm{d}\mathbf{p},
\]
is a normalization constant or the partition function. The polarity vector $\mathbf{n}$ is given by the first moment of this distribution:
\begin{equation}
\mathbf{n} = \langle \mathbf{p} \rangle = \dfrac{1}{Z} \int_{{\mathcal{H}}} \mathbf{p} \, \exp(\mathbf{b} \cdot \mathbf{p}) \, \mathrm{d}\mathbf{p}.
\end{equation}
We aim to express the higher-order moments in terms of a given polarity field $\mathbf{n}$; to this end, it is necessary to determine the relationship between $\mathbf{b}$ and $\mathbf{n}$. By symmetry, the polarity vector must be collinear with $\mathbf{b}$. We therefore work in a rotated coordinate frame in which $\mathbf{n} = |\mathbf{n}| \hat{\mathbf{z}}$, implying $\mathbf{b} = b \hat{\mathbf{z}}$. In this frame, the magnitude of the polarity is related to the scalar parameter $b$ through
\begin{equation}\label{eq:vMF relation}
|\mathbf{n}| = 1 - \frac{1}{b} + \frac{1}{e^b - 1}.
\end{equation}
The above mapping, in principle, allows us to determine $b$ for a given $\bn$, and the higher-order moments can then be approximated accordingly. However, Eq.~\eqref{eq:vMF relation} provides $|\mathbf{n}|$ as a function of $b$ and the closure requires the inverse mapping $b(|\mathbf{n}|)$. This inversion involves solving a transcendental equation at every point of the domain, and performing it repeatedly during time evolution is computationally prohibitive. 

To circumvent this bottleneck, we exploit the fact that both $|\mathbf{n}|$ and all higher-order moments depend solely on the single scalar parameter $b$; consequently, the higher-order moments can be expressed as functions of $|\mathbf{n}|$. We precompute these relationships prior to our simulations and approximate their mapping using an 8th-order Chebyshev expansion, which provides an accurate and efficient representation. This approach allows us to evaluate the higher-order moments directly from $|\mathbf{n}|$ at each time step without solving for $b$ explicitly. Once the higher-order moments are obtained in the aligned frame, they are appropriately rotated back to the lab frame, as outlined in \cite{WEADY2022110937}.

\subsection{Boundary integral method}

With the orientational moments and active forces specified through the Bingham closure, the problem reduces to solving the forced Stokes equations to determine the resulting fluid–structure interactions. A direct volumetric discretization of the Stokes equations, subject to boundary conditions and the force-free and torque-free constraints, would be computationally expensive. Instead, we employ the Boundary Integral Method (BIM) \cite{kim2005microhydrodynamics} that recasts the governing PDEs as integral equations on the surfaces. In this approach, we represent the fluid velocity at any point $\mathbf{x}$ in the domain as a combination of two single-layer operators given as
\begin{equation}\label{eq:BIE for del D and del S}
    \mathbf{u}(\mathbf{x}) = -\int_{\partial D} \mathbf{G}(\bx,\by) \cdot \mathbf{f}^w(\mathbf{y}) \, \mathrm{d}A - \int_{\partial S} \mathbf{G}(\bx,\by) \cdot \mathbf{F}_a(\mathbf{y}) \, \mathrm{d}A.
\end{equation}
Here $\mathbf{f}^w(\mathbf{y})$ is the unknown hydrodynamic traction exerted by the fluid on the rigid-body surface $\partial D$, and $\mathbf{F}_a(\mathbf{y})$ is the known active traction jump on the offset surface $\partial S$ (see Fig.~\ref{fig:rod_approximation}-$(b)$). The kernel $\mathbf{G}(\bx,\by)$ is the 3D free-space Green’s function for Stokes flow,
\begin{equation}
    \mathbf{G}(\bx,\by)=\dfrac{1}{8\pi} \left( \dfrac{\mathbf{I}}{\|\mathbf{x} - \mathbf{y}\|} + \dfrac{(\mathbf{x} - \mathbf{y})(\mathbf{x} - \mathbf{y})}{\|\mathbf{x} - \mathbf{y}\|^3} \right).
\end{equation}
To determine the unknown traction $\mathbf{f}^w(\by)$, we impose the no-slip boundary condition on the rigid surface $\partial D$,
\begin{equation}
\mathbf{u}(\mathbf{y}) = \mathbf{U} + \boldsymbol{\Omega} \times (\mathbf{y} - \mathbf{x}_0).
\end{equation}
The integrals in Eq.~\eqref{eq:BIE for del D and del S} are discretized over the rigid surface $\partial D$ and the interface $\partial S$ using a fifth-order accurate quadrature scheme. To maintain accuracy for evaluating interactions between quadrature points that are placed close to one another, we employ Quadrature by Expansion (QBX)  \cite{rachh2017fast,corona2017integral}. For the simulations presented here, the surface is resolved using approximately $N \approx 1600$ quadrature points. The resulting discretization yields the linear system
\begin{equation} \label{eq:no-slip}
\mathbf{U} + \boldsymbol{\Omega} \times (\mathbf{y} - \mathbf{x}_0)
= -\mathbf{S}_{\partial D \rightarrow \partial D} \cdot \mathbf{f}^w
- \mathbf{S}_{\partial S \rightarrow \partial D} \cdot \mathbf{F}^a,
\end{equation}
where $\mathbf{S}_{\partial D \rightarrow \partial D}$ denotes the discrete single-layer operator mapping tractions on $\partial D$ to induced velocities on $\partial D$, and $\mathbf{S}_{\partial S \rightarrow \partial D}$ maps the traction jump on $\partial S$ to velocities on $\partial D$. The force-free and torque-free constraints are discretized as
\begin{align}
\mathbf{W} \cdot \mathbf{f}^w
&= -\mathbf{W} \cdot \mathbf{I}_n \cdot \mathbf{F}_a, \\
\boldsymbol{\zeta} \cdot \mathbf{f}^w
&= -\boldsymbol{\zeta} \cdot \mathbf{I}_n \cdot \mathbf{F}_a
- \mathbf{W} \cdot \langle \mathbf{T}_0 \rangle ,
\end{align}
where $\mathbf{W}$ contains the quadrature weights associated with integration over $\partial D$, while $\boldsymbol{\zeta}$ encodes the weighted moment arms such that
$\boldsymbol{\zeta} \cdot \mathbf{f}^w = \int_{\partial D} (\mathbf{y}-\mathbf{x}_0) \times \mathbf{f}^w \, \mathrm{d}A$. The interpolation matrix $\mathbf{I}_n$ maps quantities defined on the interface $\partial S$ to the corresponding quadrature points on $\partial D$.

The discretized equations are assembled into a single linear system for the unknown surface tractions and rigid-body velocities,
\begin{align} \left[\begin{array}{c|c|c} \mathbf{S}_{\partial D \rightarrow \partial D} & \mathbf{I}_d & \mathbf{X} \\ \hline \mathbf{W} & \mathbf{O} & \mathbf{O} \\ \hline \boldsymbol{\zeta} & \mathbf{O} & \mathbf{O} \end{array}\right] \begin{bmatrix} \mathbf{f}^w \\ \mathbf{U} \\ \boldsymbol{\Omega} \end{bmatrix} = \begin{bmatrix} -\mathbf{S}_{\partial S \rightarrow \partial D} \cdot \mathbf{F}_a \\ -\mathbf{W}\cdot \mathbf{I}_n \cdot \mathbf{F}_a \\ -\boldsymbol{\zeta}\cdot \mathbf{I}_n \cdot \mathbf{F}_a - \mathbf{W} \cdot \langle\bT_0\rangle \end{bmatrix}, \label{eq:Lin. eqn} \end{align}
which is solved simultaneously for $\mathbf{f}^w$, $\mathbf{U}$, and $\boldsymbol{\Omega}$. Here, $\mathbf{I}_d$ consists of stacked $3\times3$ identity blocks, $\mathbf{X}$ encodes the geometric cross-product operator associated with rigid-body rotation, and $\mathbf{O}$ denotes zero matrices of appropriate size. For a given polarity field $\mathbf{n}$, Eq.~\eqref{eq:Lin. eqn} determines the wall tractions $\mathbf{f}^w$ and the rigid-body velocities $\{\mathbf{U},\boldsymbol{\Omega}\}$. 

Note that the traction jump $\mathbf{F}^a$ depends on the surface velocity gradient $\nabla \mathbf{u}|_{\partial D}$, which itself depends on the solution of the Stokes problem. We account for this coupling by computing $\nabla \mathbf{u}|_{\partial D}$ using an implicit scheme. Exploiting the linearity of the Stokes equations, the dependence of $\nabla \mathbf{u}|_{\partial D}$ on the unknown tractions can be recast as a linear problem for the surface velocity gradients. Following the approach of Ref.~\cite{doi:10.1073/pnas.2405114121}, we solve this system iteratively using the Generalized Minimal Residual (GMRES) method, which converges within 15–20 iterations. The temporal evolution of the system is then done using a second-order accurate Runge–Kutta (RK2) scheme with time step $\Delta t \sim 10^{-2}$. All boundary integral operators in our numerical implementation are evaluated using the \texttt{FMM3DBIE} library, an open-source fast multipole accelerated boundary integral solver for Stokes flow \cite{rachh2017fast}.

\section{Results: nonlinear simulations}\label{sec:nonlin}
\subsection{The fixed sphere: swirling instability}\label{subsec:swirling}
\begin{figure}
    \centering
    \includegraphics[width=1\linewidth]{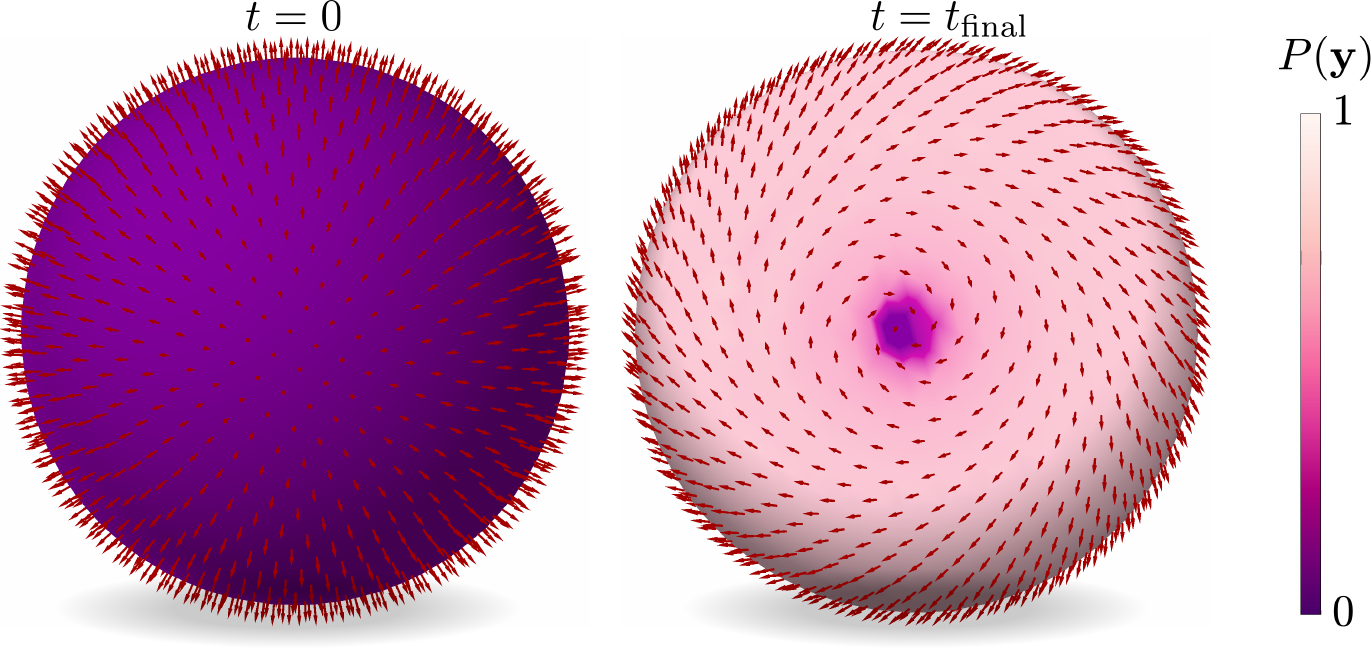}
    \caption{Evolution of the polarity field on a fixed sphere with uniform nucleation.
    Snapshots of the polarity field during its evolution from the initial state ($t=0$, left) to the final steady state ($t=t_{\mathrm{final}}$, right). The initial configuration is nearly radial, whereas the steady state develops a swirling pattern containing two antipodal $+1$ topological defects. Away from these defects, the filaments maintain an approximately constant tilt angle relative to the local surface normal, generating a coherent rotational flow. Arrows indicate the polarity orientation, and the color map shows the tangential order parameter $P(\mathbf{y}) = \left\| \left(\mathbf{I} - \hat{\mathbf{e}}_r \hat{\mathbf{e}}_r \right)\cdot \mathbf{n}(\mathbf{y}) \right\|$.
    }
    \label{fig:fixed_sphere}
\end{figure}
We start our investigation of the nonlinear dynamics by simulating a fixed rigid sphere ($\mathbf{U}=\mathbf{0},  \boldsymbol{\Omega}=\mathbf{0}$) coated with a uniform distribution of nucleating filaments. In analogy with the half-space problem, this configuration also admits a linearly unstable fixed point where all filaments point radially outward, ${\bn}=\hat{\mathbf{e}}_r$, and the mean-field flow vanishes, $\bu=\mathbf{0}$. We initialize the simulations with a small perturbation around this state. As the instability develops, correlated regions of high polar order emerge spontaneously among neighboring filaments, accompanied by the formation of multiple topological defects on the spherical surface.
\begin{figure*}
    \centering
    \includegraphics[width=1\linewidth]{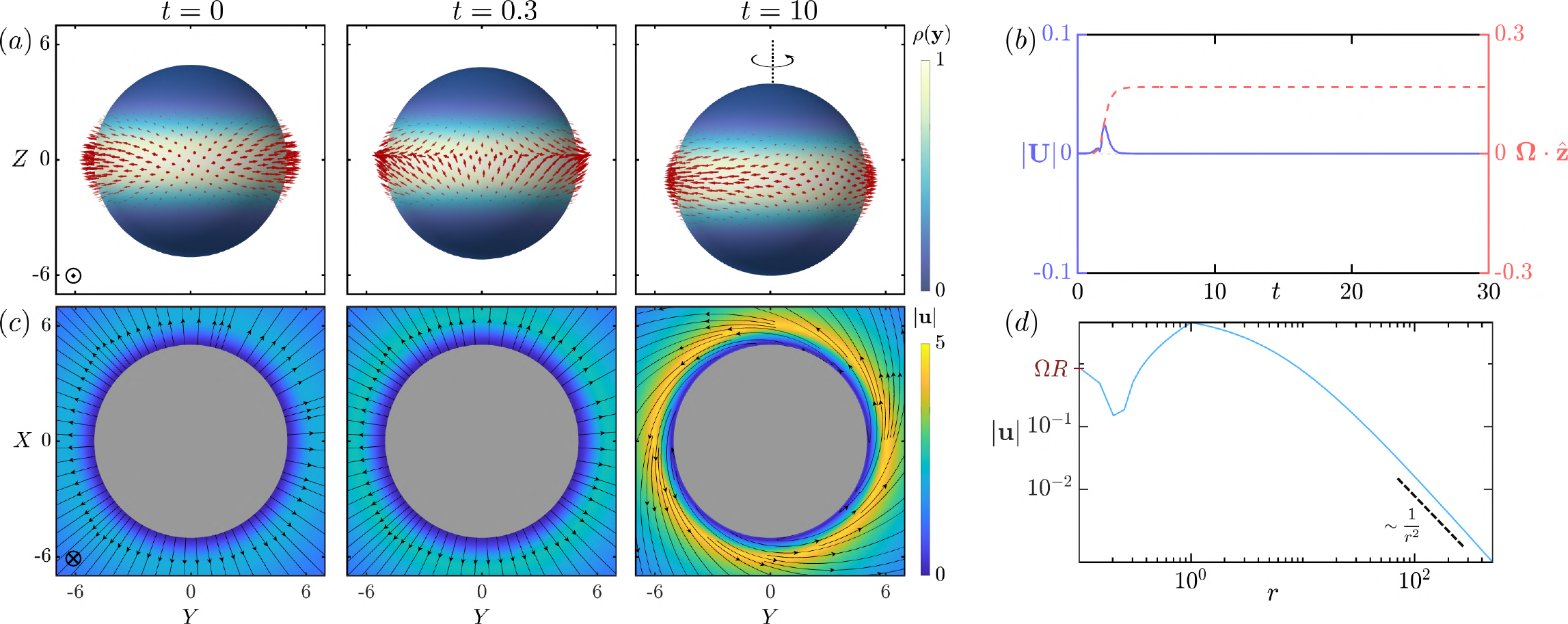}
    \caption{Dynamics and flow field in the equatorial band configuration.
    (a) Snapshots of the polarity field $\mathbf{n}(\mathbf{y})$ on the spherical surface at times $t=0$, $0.3$, and $10$. The surface color indicates the filament density $\rho(\mathbf{y})$ (color bar, right), while arrows denote the polarity orientation. Initially, the polarity is aligned with the local surface normal; at later times, it reorients along the imposed nucleation pattern, leading to sustained rotation of the rigid body.
    (b) Time evolution of the translational speed $|\mathbf{U}|$ (blue) and the angular velocity component $\boldsymbol{\Omega}\cdot\hat{\mathbf{z}}$ (red). After a transient increase in translation, the translational velocity decays to zero, while the angular velocity approaches a nonzero steady value corresponding to persistent rotation.
    (c) Flow field in the $z=0$ plane; the color indicates the flow magnitude $|\mathbf{u}|$. The initial flow is weak and predominantly radial, whereas the steady-state flow develops a strong tangential component associated with body rotation. Near the surface, the tangential flow follows the direction of rotation, while farther from the body, it reverses direction.
    (d) Radial decay of the flow speed, showing the far-field scaling $|\mathbf{u}| \sim r^{-2}$.}
    \label{fig:theta-patch}
\end{figure*}
The defects are characterized using a surface order parameter $P(\by) = \| (\mathbf{I} - \hat{\mathbf{e}}_r \hat{\mathbf{e}}_r) \cdot \mathbf{n}(\by) \|$ which measures the magnitude of the tangential component of the filament polarity. This quantity takes values $P\approx 0$ near defect cores and $P\approx 1$ in regions where filaments are strongly aligned. As the dynamics proceed, defects migrate and merge, and the system relaxes to a steady state containing two $+1$ defects located at the opposite poles of the sphere, consistent with the Poincaré–Hopf theorem (see Fig.~\ref{fig:fixed_sphere}-$(b)$). This resulting global polarity pattern bears a striking resemblance to the organization underlying the internal cytoplasmic twister flows observed in \textit{Drosophila} oocytes \cite{Dutta2024,doi:10.1073/pnas.2405114121,PRXLife.3.023007,PhysRevLett.126.028103} where molecular motors walking along cortical microtubules generate tangential forces that drive large-scale vortical flows in the cell \cite{glotzer1997cytoplasmic,quinlan2016cytoplasmic,ganguly2012cytoplasmic,khuc2015cortical}. In our system, filament polymerization at the surface similarly generates tangential forces on the surrounding fluid, producing similar coherent rotational flows exterior to the sphere.

\subsection{Free sphere: spontaneous rotation}

We now allow the sphere to translate and rotate freely in response to the hydrodynamic forces generated by the filaments. Starting from the same nearly radial polarity field considered in the fixed-sphere case, the instability again develops, and the surface polarity organizes into the two-defect swirling configuration described in the previous section. This symmetry-broken surface flow now produces a net hydrodynamic torque on the body. As a result, conservation of angular momentum demands that the sphere undergo a sustained rigid-body rotation about a spontaneously selected axis. It is interesting to note that this observed dynamics is independent of initial conditions on the polarity field. We observed that certain initial conditions can lead to transient translation of the sphere; however, the sphere eventually settles to a state of spinning whereby the active torques from polymerizing filaments are balanced by the drag forces on the spinning sphere.

\subsection{Patterned nucleation and motility}

In the uniformly coated sphere, activity produces a torque but no net force, leading only to rigid-body rotation. This raises a natural question: can breaking the spatial symmetry of filament density generate directed motion of the body? Here, we explore this by considering patterned nucleation sites on the surface.

We first examine an equatorial band on a sphere, defined by a Gaussian density profile of nucleation sites centered on the equator (see Fig.~\ref{fig:theta-patch}-$(a)$). The system is initialized with all filaments oriented along their local surface normals and the rigid body at rest. This state is generically unstable, as the filaments bend and the ensuing hydrodynamic interactions cause them to tilt collectively away from the radially outwards configuration. During this transient stage, the asymmetric traction distribution across the equator generates a non-zero translational velocity (see Fig.~\ref{fig:theta-patch}-$(b)$) and an associated translation of the sphere, as can be seen from Fig.~\ref{fig:theta-patch}-$(a)$.

As the dynamics proceed, the filaments progressively align tangentially along the equator, organizing into a coherent band around the equator (see Fig.~\ref{fig:theta-patch}-$(a)$, right panel). Because filament nucleation is localized near the equator, the polarity field is defined only over this band and can be devoid of topological defects, as seen in the steady state: a behavior which contrasts with that of the uniformly coated sphere. The progressive realignment into the equatorial direction leads to the loss of sphere translation and emergence of sphere rotation, as shown in Fig.~\ref{fig:theta-patch}-$(b)$. The associated flow field in the equatorial plane of the sphere is shown in Fig.~\ref{fig:theta-patch}-$(c)$. The fluid velocity in the ultimate steady state is maximal within the narrow traction layer surrounding the sphere, where the active forces are applied, and decays away from the surface; it is interesting to highlight that the steady-state flow is not entirely azimuthal, as the buckled filaments splay out away from the equator, generating weak radial components that result in a spiraling flow. In the far field, the velocity exhibits a $|\bu|\sim r^{-2}$ decay, consistent with the hydrodynamic signature of neutrally buoyant force-free microswimmers \cite{Lauga_2009}.

\begin{figure}
    \centering
    \includegraphics[width=1\linewidth]{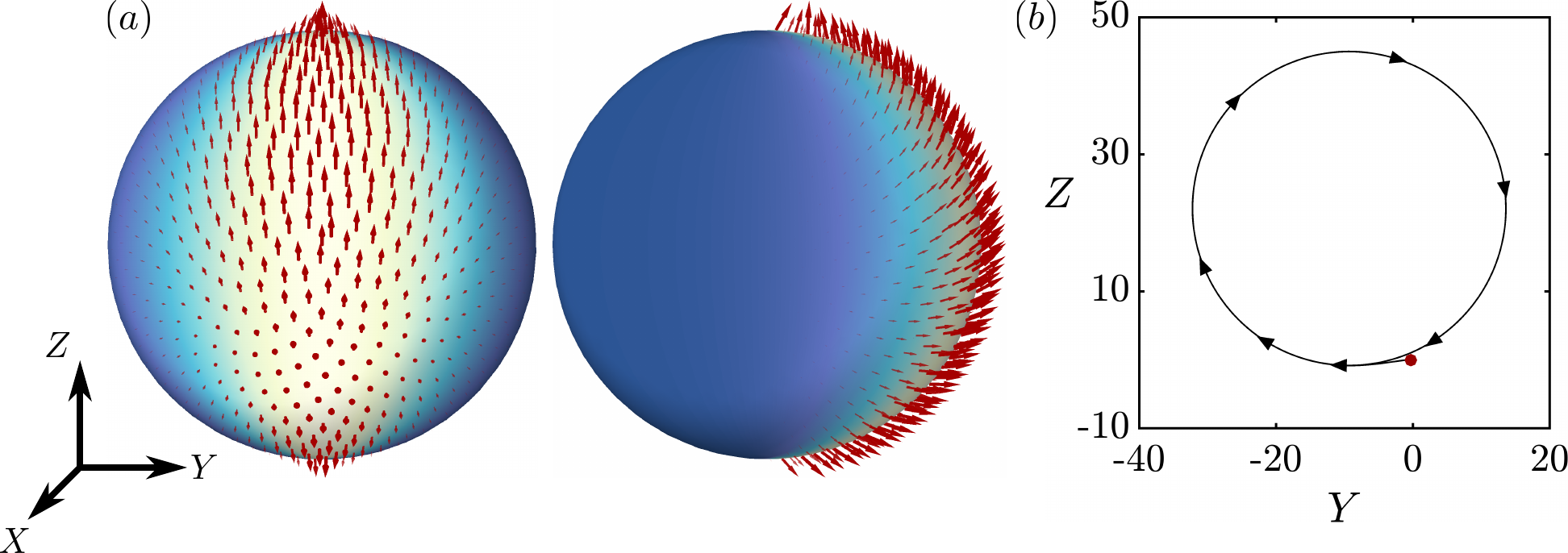}
    \caption{Dynamics of the sphere in the polar-patch configuration. (a) Steady-state polarity field on the spherical surface for the polar-patch configuration. The arrows indicate the polarity orientation, which tilts in the direction set by the imposed filament distribution. The surface color represents the filament density $\rho(\mathbf{y})$.
    (b) Trajectory of the sphere’s center of mass, showing sustained circular motion in the $YZ$-plane. The red dot denotes the initial position.}
    \label{fig:phi-patch}
\end{figure}

Next, we consider the \textit{polar patch}, which is the natural dual of the equatorial band discussed above. In this case, the filament density is localized in the azimuthal direction and is prescribed by a Gaussian profile centered at the azimuthal angle $\phi = \phi_0$, where $\phi \in [0,2\pi)$ denotes the azimuthal coordinate around the $z$-axis; without loss of generality, here we take $\phi_0=\pi$. The system evolves to a steady configuration in which the filaments again tilt by a constant angle relative to their local surface normals and align along the direction of the filament distribution itself (Fig.~\ref{fig:phi-patch}-$(a)$). However, in contrast to the equatorial band, the symmetry of the traction distribution no longer eliminates the net force on the body. As a result, the rigid sphere exhibits both rotation and translation.
The origin of this combined motion can be understood by considering two limiting configurations of the filament orientation. First, we note that the nucleation patch is centered at the azimuthal location $\phi_0=\pi$; if the growing filaments had remained aligned with their local surface normals, the active forces pointing radially outward from the patch would result in a net force directed along the $\hat{\mathbf{x}}$ axis, giving rise to pure translation of the sphere. In contrast, if the filaments buckle and align tangentially along the $\hat{\theta}$ direction, the resulting associated traction generates both a net force along $-\hat{\mathbf{z}}$ and a constant torque about the $x$-axis. The emergent steady state realized in the simulations corresponds to an intermediate configuration between these idealized limits, producing simultaneous translation and rotation. Consequently, the sphere moves along a closed circular trajectory while maintaining a constant angular velocity, as shown in Fig.~\ref{fig:phi-patch}-$(b)$.

\begin{figure}[H]
    \centering
    \includegraphics[width=1\linewidth]{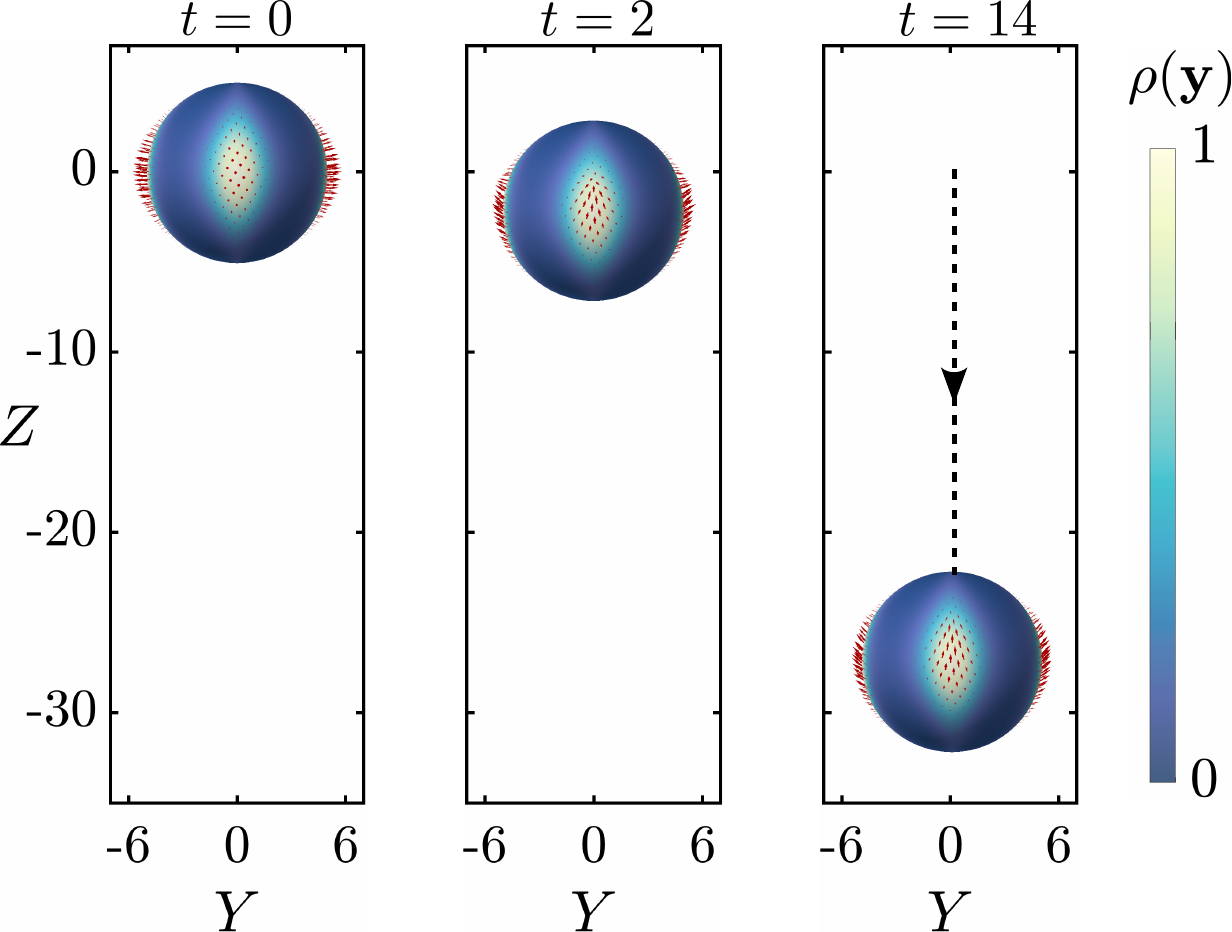}
    \caption{Time evolution of the polarity field on the sphere's surface in the four-sided configuration.
    Initially ($t=0$), the polarity is aligned with the surface normal. At later times ($t=2$ and $t=14$), the polarity tilts upward, resulting in a steady downward translation of the rigid body. The dashed line represents the trajectory of the rigid body's centre of mass.}
    \label{fig:4fold}
\end{figure}

The patterned nucleation sites explored above do not, by themselves, produce a persistent swimmer, but they provide valuable insight into how surface patterning controls the balance of forces and torques on the body. In particular, the equatorial band generates a net torque but no net force, leading to pure spinning of the sphere. In contrast, the polar patch produces both force and torque, resulting in circular trajectories. These observations suggest that by exploiting symmetry in the nucleation pattern, one can cancel the torques generated by individual patches while allowing their force contributions to add constructively, thereby producing net propulsion. Motivated by this idea, we construct a nucleation pattern consisting of four identical patches arranged with four-fold symmetry around the sphere. The patches are centered at azimuthal locations $\phi_j = j\pi/2$, where $j=1,\ldots,4$, as illustrated in Fig.~\ref{fig:4fold}. In this configuration, the torques generated by individual patches cancel by symmetry, while their force contributions add constructively. As a result, the sphere reaches a steady state characterized by persistent translation along the $-z$ direction in the YZ-plane, as shown in Fig.~\ref{fig:4fold}.

\begin{figure}
    \centering
    \includegraphics[width=1\linewidth]{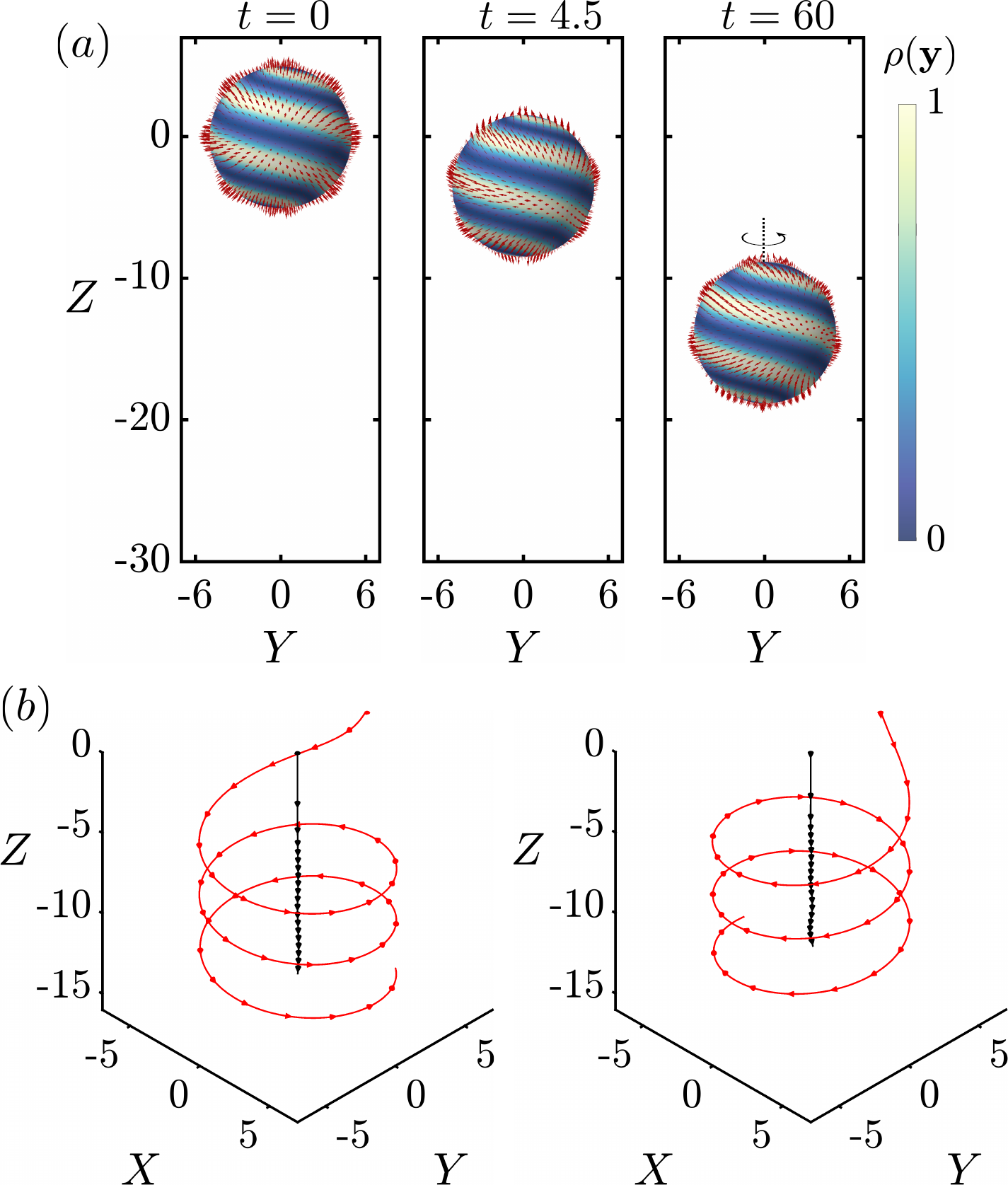}
    \caption{(a) Snapshots of the polarity field $\bn(\by)$ on the sphere in the chiral configuration. At $t = 0$ the polarity is axisymmetric; by $t = 4.5$ and $t = 60$ it has tilted to align with the helical nucleation pattern, breaking the axial symmetry and driving simultaneous translation and rotation of the body. (b) Trajectories of the sphere’s center of mass (black) and a reference point on its surface (red) for chiral configurations with opposite handedness. The left and right panels correspond to left- and right-handed spiral nucleation, respectively. In both cases, the center of mass follows a straight path, while the surface point traces helical trajectories of opposite handedness.}
    \label{fig:spiral_patch}
\end{figure}

While many other surface patterns could, in principle, lead to propulsion, a particularly intriguing possibility is inspired by chirality-driven propulsion mechanisms observed in nature. In bacterial motility, for example, propulsion arises from the rotation of a helical flagellum, which converts rotational motion into translation through the chiral coupling inherent to the helix. Motivated by this idea, we consider a chiral nucleation pattern in which filament growth sites are arranged along a helical band on the surface of the particle. Following the onset of the instability, the filaments buckle and align along the direction of the nucleation sites, producing a helical distribution of traction on the particle surface. This chiral traction pattern generates a translation–rotation coupling: the particle spins about the symmetry axis of the helix while simultaneously translating along it (see Fig.~\ref{fig:spiral_patch}). Similar chiral propulsion strategies are found in the plant sperm of \textit{Ginkgo biloba}, where thousands of flagella arranged in a helical band on the frustum-shaped cell body generate metachronal waves that drive the cell forward \cite{Li1989}. In our system, the helical arrangement of the nucleation sites plays an analogous role, directly encoding chirality into the surface pattern. As a consequence, the direction of propulsion and the handedness of the resulting trajectory can be programmed through the handedness of the nucleation pattern itself (see Fig.~\ref{fig:spiral_patch}-$(b)$).

\subsection{Role of particle geometry}

One of the biological examples motivating the present work is the motility of \textit{Acetobacter xylinum} whose cell body is closer to a spheroid. Since particle geometry plays a central role in determining both the hydrodynamic response and the ensuing dynamics of the microswimmer, we extend our analysis from spherical to spheroidal bodies to examine how shape influences the emergent dynamics driven by filament activity.

\begin{figure}[H]
    \centering
    \includegraphics[width=1\linewidth]{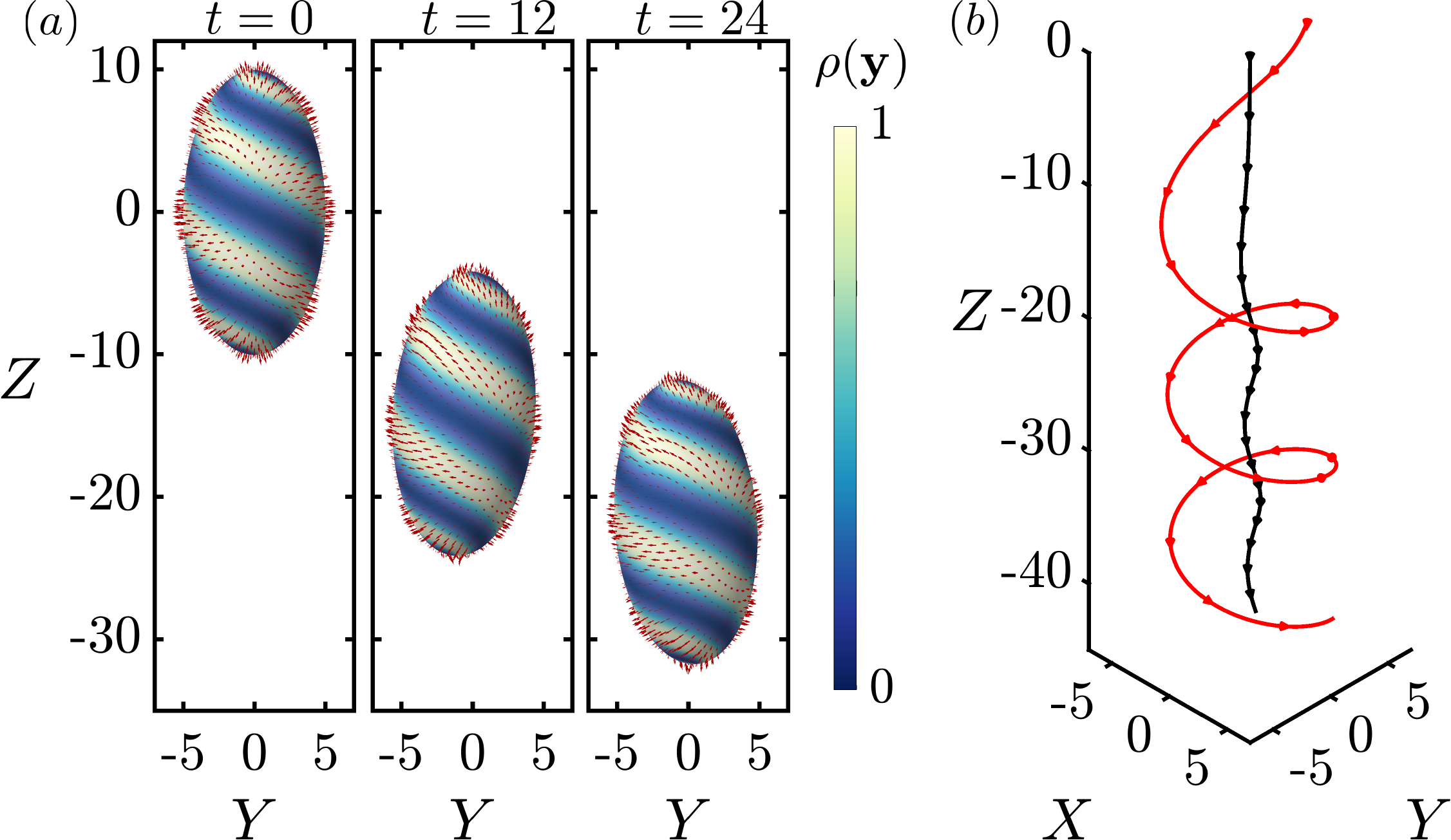}
    \caption{%
        \textbf{(a)}~Snapshots of the polarity field $\bn(\mathbf{y})$ on the prolate spheroid in the chiral configuration. As before, initially the polarity is axisymmetric and eventually breaks the axial symmetry, driving
        simultaneous translation and rotation of the body.
        \textbf{(b)}~Trajectories of the spheroid's center of mass (black) and of a reference point on its surface (red). In contrast to the spherical case, the center of mass traces a tightly wound circular helix, reflecting the continuous precession of the propulsion direction induced by the residual polar asymmetry and the drag anisotropy of the spheroidal geometry.%
    }
    \label{fig:spiral_spheroid}
\end{figure}

To this end, we consider a prolate spheroid with a chiral arrangement of nucleation sites. As in the spherical case, the steady-state polarity field tilts to align with the filament distribution (Fig.~\ref{fig:spiral_spheroid}-$(a)$). The resulting dynamics, however, differ qualitatively from the spherical case: rather than translating along a straight line, the center of mass of the spheroid traces a circular helix with a
small but finite radius (Fig.~\ref{fig:spiral_spheroid}-$(b)$). Circular helical trajectories have also been reported for active swimmers with prescribed chiral surface slip velocity~\cite{das2026squirmersarbitraryshapeslip}; however, the mechanism in the present model is fundamentally different.

Here, the helical trajectory arises from an emergent weak \emph{polar asymmetry} in the steady-state polarity field, whereby the net traction on the body is not perfectly aligned with the principal axis of the spheroid. This misalignment causes the propulsion direction to precess continuously, generating the observed helical path. Although a small polar asymmetry is also present in the spherical geometry, its dynamical consequence is more pronounced for the spheroid, where the anisotropic Stokes drag couples the misaligned force more strongly to rotational motion.

\section{Discussion}

The central contribution of this work is a self-consistent continuum framework that connects microscopic polymerization kinetics to macroscopic swimming through the collective behavior of a surface-anchored filament carpet. In this model, the active forces emerge self-consistently from filament growth, orientational dynamics, and hydrodynamic interactions. Our results demonstrate that the spatial pattern of nucleation sites encodes the balance of forces and torques on the 
body: an equatorial band drives pure spinning, a fourfold-symmetric arrangement 
yields directed translation, and a chiral pattern couples rotation to translation. Extending the analysis to spheroidal geometries further reveals that body shape qualitatively alters the swimming trajectory through drag anisotropy.

\begin{figure}[H]
    \centering
    \includegraphics[width=0.6\linewidth]{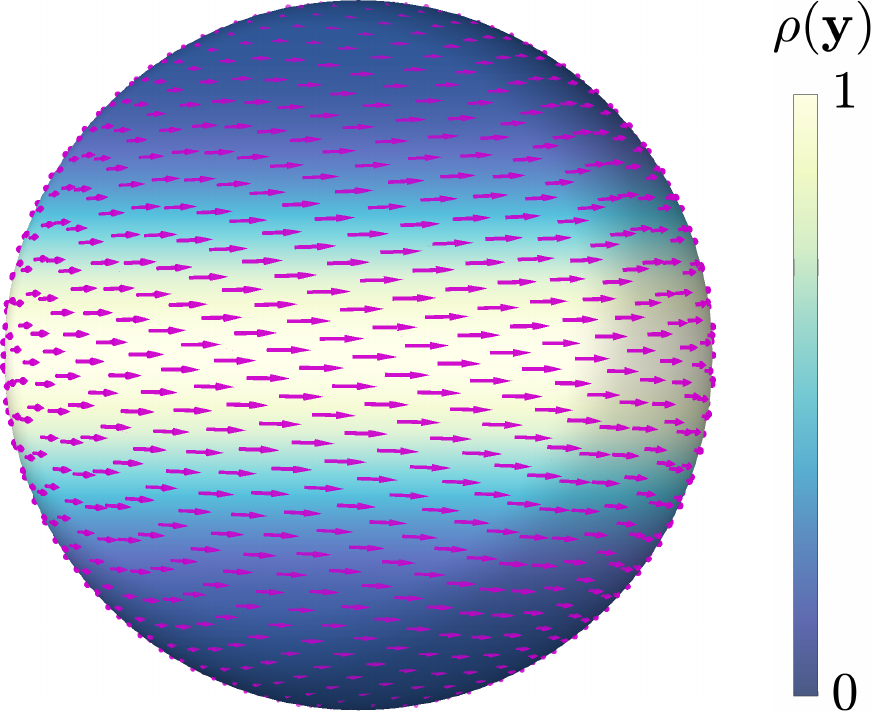}
    \caption{Slip velocity induced by the active carpet in the equatorial band configuration.
    The velocity field computed on the effective surface $\partial S$ is imposed as a slip velocity $\mathbf{v}_s$ on the rigid body, mapping the system onto the standard squirmer model. Arrows indicate the direction and magnitude of the slip velocity field $\mathbf{v}_s$. The color map shows the filament density ${\rho}(\mathbf{y})$ on the body surface $\partial D$.}
    \label{fig:squirmer}
\end{figure}

A natural reference point for any microswimmer model is the classical squirmer 
\cite{Lighthill1952, blake1971spherical}, which represents the effect of surface 
activity as a prescribed tangential slip velocity, with the resulting rigid-body 
motion computed via the reciprocal theorem~\cite{PhysRevLett.77.4102}:
\begin{align}
    \mathbf{U} &= -\dfrac{1}{4\pi} \iint_{\partial D} \mathbf{v}_s \,\mathrm{d}S, \\
    \boldsymbol{\Omega} &= -\dfrac{3}{8\pi} \iint_{\partial D} \hat{\mathbf{n}} 
    \times \mathbf{v}_s \,\mathrm{d}S,
\end{align}
where $\mathbf{v}_s$ is a surface slip velocity. Mapping our filament-resolved dynamics to an equivalent squirmer description by identifying the tangential fluid velocity on the effective traction surface $\partial S$ as a slip velocity $\mathbf{v}_s$ on $\partial D$ (Fig.~\ref{fig:squirmer}) reveals a systematic discrepancy. The squirmer prediction overestimates the steady angular velocity by a factor of approximately $1.5$ relative to the full model: a quantitative mismatch that possibly stems from the ambiguity in choosing an appropriate surface for computing slip velocity.

The micro--macro framework developed here is of broader relevance beyond polymerization-driven motility. The same modeling architecture involving adiabatic elimination of fast filament-length dynamics, Bingham closure for orientational moments, and a traction-layer boundary integral formulation applies directly to motor-driven systems where anchored filaments are reoriented by molecular motors rather than by growth. The Cytoplasmic streaming problems in \textit{Drosophila} oocytes \cite{Dutta2024} and \textit{C. elegans} zygotes \cite{kimura2017endoplasmic} are examples that have been successfully analyzed with closely related methods~\cite{doi:10.1073/pnas.2405114121}. The present work extends this class of frameworks to external-fluid problems and free swimmers, suggesting that polymerization and motor activity may be treated within a unified continuum description.

The framework also suggests a natural extension to intracellular self-organization problems of a different geometry. When microtubules nucleate from a central microtubule-organizing center (MTOC) and grow radially outward, their free plus-ends can 
encounter the cell cortex and buckle under polymerization load, generating 
collective forces that position the aster — directly relevant to the spindle positioning in large cells~\cite{sulerud2020microtubule}.  This setting is, in effect, the geometric dual of the one studied here, with the domain turned inside out: our filaments radiate outward from a body into an unbounded fluid, where viscous resistance supplies the compression that drives buckling, whereas in the aster they radiate from a central MTOC toward an enclosing cortex, against which the plus-ends buckle on contact. The same growth-induced buckling reorients the filaments and sets the net force in both cases; the bounded interior geometry then results in centering, whereas the unbounded exterior, studied here, yields locomotion.

On the synthetic design side, our results demonstrate that swimming speeds of  $\mathcal{O}(0.1$--$1\ \mu\mathrm{m/s})$ are achievable purely from polymerization-driven collective dynamics. The nucleation pattern functions as a programmable surface code: its symmetry determines whether the body spins,  translates, or follows helical trajectories. This design perspective connects to recent work on light-controlled surface-driven active carpets, where topological patterning of boundary activity was shown to engineer complex bulk flow patterns in confined geometries~\cite{gong2021engineering}, and to experimental demonstrations of polymerization-propelled synthetic colloids~\cite{lopes2025emergenceantichemotacticflockingactive}.

The present work also opens several directions for future investigation. The self-organized flows and programmable trajectories demonstrated here for a single swimmer naturally raise the question of how hydrodynamic interactions between pairs or collections of polymerization-driven swimmers give rise to collective behavior — a problem we intend to pursue. A more fundamental modeling challenge is posed by the long-filament regime characteristic of  \textit{Acetobacter xylinum}, where cellulose fibers grow to lengths exceeding the bacterial body size, violating the thin-carpet assumption underpinning the 
traction-layer approximation. In this regime, inter-filament steric interactions, hydrodynamic coupling at finite separation, and large deformations of buckled fibers extending far into the fluid all demand new theoretical machinery. Extending the present framework to address these regimes — perhaps through hybrid discrete-continuum approaches remains an open and tractable challenge that the methods developed here help to motivate.

\section{Acknowledgement}

N.K.D thanks Devadevan M M and Harshit Joshi for many helpful discussions on the work. B.C. acknowledges support from the Department of Atomic Energy, Government of India (under project number RTI4019) and the ANRF ECRG grant 002268. 



\bibliographystyle{rsc}
\bibliography{bibtex}

\providecommand*{\mcitethebibliography}{\thebibliography}
\csname @ifundefined\endcsname{endmcitethebibliography}
{\let\endmcitethebibliography\endthebibliography}{}
\begin{mcitethebibliography}{51}
\providecommand*{\natexlab}[1]{#1}
\providecommand*{\mciteSetBstSublistMode}[1]{}
\providecommand*{\mciteSetBstMaxWidthForm}[2]{}
\providecommand*{\mciteBstWouldAddEndPuncttrue}
  {\def\EndOfBibitem{\unskip.}}
\providecommand*{\mciteBstWouldAddEndPunctfalse}
  {\let\EndOfBibitem\relax}
\providecommand*{\mciteSetBstMidEndSepPunct}[3]{}
\providecommand*{\mciteSetBstSublistLabelBeginEnd}[3]{}
\providecommand*{\EndOfBibitem}{}
\mciteSetBstSublistMode{f}
\mciteSetBstMaxWidthForm{subitem}
{(\emph{\alph{mcitesubitemcount}})}
\mciteSetBstSublistLabelBeginEnd{\mcitemaxwidthsubitemform\space}
{\relax}{\relax}

\bibitem[Purcell(1977)]{10.1119/1.10903}
E.~M. Purcell, \emph{American Journal of Physics}, 1977, \textbf{45},
  3--11\relax
\mciteBstWouldAddEndPuncttrue
\mciteSetBstMidEndSepPunct{\mcitedefaultmidpunct}
{\mcitedefaultendpunct}{\mcitedefaultseppunct}\relax
\EndOfBibitem
\bibitem[Batchelor(2000)]{Batchelor_2000}
G.~K. Batchelor, in \emph{Flow of a Uniform Incompressible Viscous Fluid},
  Cambridge University Press, 2000, p. 174–263\relax
\mciteBstWouldAddEndPuncttrue
\mciteSetBstMidEndSepPunct{\mcitedefaultmidpunct}
{\mcitedefaultendpunct}{\mcitedefaultseppunct}\relax
\EndOfBibitem
\bibitem[Berg(2003)]{berg2003rotary}
H.~C. Berg, \emph{Annual Review of Biochemistry}, 2003, \textbf{72},
  19--54\relax
\mciteBstWouldAddEndPuncttrue
\mciteSetBstMidEndSepPunct{\mcitedefaultmidpunct}
{\mcitedefaultendpunct}{\mcitedefaultseppunct}\relax
\EndOfBibitem
\bibitem[Taylor(1951)]{doi:10.1098/rspa.1951.0218}
G.~I. Taylor, \emph{Proceedings of the Royal Society of London. Series A.
  Mathematical and Physical Sciences}, 1951, \textbf{209}, 447--461\relax
\mciteBstWouldAddEndPuncttrue
\mciteSetBstMidEndSepPunct{\mcitedefaultmidpunct}
{\mcitedefaultendpunct}{\mcitedefaultseppunct}\relax
\EndOfBibitem
\bibitem[Elgeti and Gompper(2013)]{doi:10.1073/pnas.1218869110}
J.~Elgeti and G.~Gompper, \emph{Proceedings of the National Academy of
  Sciences}, 2013, \textbf{110}, 4470--4475\relax
\mciteBstWouldAddEndPuncttrue
\mciteSetBstMidEndSepPunct{\mcitedefaultmidpunct}
{\mcitedefaultendpunct}{\mcitedefaultseppunct}\relax
\EndOfBibitem
\bibitem[Machemer(1972)]{Machemer1972}
H.~Machemer, \emph{Journal of Experimental Biology}, 1972, \textbf{57},
  239--259\relax
\mciteBstWouldAddEndPuncttrue
\mciteSetBstMidEndSepPunct{\mcitedefaultmidpunct}
{\mcitedefaultendpunct}{\mcitedefaultseppunct}\relax
\EndOfBibitem
\bibitem[Aoun \emph{et~al.}(2020)Aoun, Farutin, Garcia-Seyda, Nègre, Rizvi,
  Tlili, Song, Luo, Biarnes-Pelicot, Galland, Sibarita, Michelot, Hivroz,
  Rafai, Valignat, Misbah, and Theodoly]{AOUN20201157}
L.~Aoun, A.~Farutin, N.~Garcia-Seyda, P.~Nègre, M.~S. Rizvi, S.~Tlili,
  S.~Song, X.~Luo, M.~Biarnes-Pelicot, R.~Galland, J.-B. Sibarita, A.~Michelot,
  C.~Hivroz, S.~Rafai, M.-P. Valignat, C.~Misbah and O.~Theodoly,
  \emph{Biophysical Journal}, 2020, \textbf{119}, 1157--1177\relax
\mciteBstWouldAddEndPuncttrue
\mciteSetBstMidEndSepPunct{\mcitedefaultmidpunct}
{\mcitedefaultendpunct}{\mcitedefaultseppunct}\relax
\EndOfBibitem
\bibitem[Farutin \emph{et~al.}(2013)Farutin, Rafa\"{\i}, Dysthe, Duperray,
  Peyla, and Misbah]{PhysRevLett.111.228102}
A.~Farutin, S.~Rafa\"{\i}, D.~K. Dysthe, A.~Duperray, P.~Peyla and C.~Misbah,
  \emph{Phys. Rev. Lett.}, 2013, \textbf{111}, 228102\relax
\mciteBstWouldAddEndPuncttrue
\mciteSetBstMidEndSepPunct{\mcitedefaultmidpunct}
{\mcitedefaultendpunct}{\mcitedefaultseppunct}\relax
\EndOfBibitem
\bibitem[Lighthill(1952)]{Lighthill1952}
M.~J. Lighthill, \emph{Communications on Pure and Applied Mathematics}, 1952,
  \textbf{5}, 109--118\relax
\mciteBstWouldAddEndPuncttrue
\mciteSetBstMidEndSepPunct{\mcitedefaultmidpunct}
{\mcitedefaultendpunct}{\mcitedefaultseppunct}\relax
\EndOfBibitem
\bibitem[Blake(1971)]{blake1971spherical}
J.~R. Blake, \emph{Journal of Fluid Mechanics}, 1971, \textbf{46},
  199--208\relax
\mciteBstWouldAddEndPuncttrue
\mciteSetBstMidEndSepPunct{\mcitedefaultmidpunct}
{\mcitedefaultendpunct}{\mcitedefaultseppunct}\relax
\EndOfBibitem
\bibitem[Ishikawa(2024)]{annurev:/content/journals/10.1146/annurev-fluid-121021-042929}
T.~Ishikawa, \emph{Annual Review of Fluid Mechanics}, 2024, \textbf{56},
  119--145\relax
\mciteBstWouldAddEndPuncttrue
\mciteSetBstMidEndSepPunct{\mcitedefaultmidpunct}
{\mcitedefaultendpunct}{\mcitedefaultseppunct}\relax
\EndOfBibitem
\bibitem[Ishikawa \emph{et~al.}(2020)Ishikawa, Pedley, Drescher, and
  Goldstein]{Ishikawa_Pedley_Drescher_Goldstein_2020}
T.~Ishikawa, T.~J. Pedley, K.~Drescher and R.~E. Goldstein, \emph{Journal of
  Fluid Mechanics}, 2020, \textbf{903}, A11\relax
\mciteBstWouldAddEndPuncttrue
\mciteSetBstMidEndSepPunct{\mcitedefaultmidpunct}
{\mcitedefaultendpunct}{\mcitedefaultseppunct}\relax
\EndOfBibitem
\bibitem[Chamolly and Ishikawa(2026)]{Chamolly_Ishikawa_2026}
A.~Chamolly and T.~Ishikawa, \emph{Journal of Fluid Mechanics}, 2026,
  \textbf{1030}, A22\relax
\mciteBstWouldAddEndPuncttrue
\mciteSetBstMidEndSepPunct{\mcitedefaultmidpunct}
{\mcitedefaultendpunct}{\mcitedefaultseppunct}\relax
\EndOfBibitem
\bibitem[Chakrabarti \emph{et~al.}(2022)Chakrabarti, Fürthauer, and
  Shelley]{doi:10.1073/pnas.2113539119}
B.~Chakrabarti, S.~Fürthauer and M.~J. Shelley, \emph{Proceedings of the
  National Academy of Sciences}, 2022, \textbf{119}, e2113539119\relax
\mciteBstWouldAddEndPuncttrue
\mciteSetBstMidEndSepPunct{\mcitedefaultmidpunct}
{\mcitedefaultendpunct}{\mcitedefaultseppunct}\relax
\EndOfBibitem
\bibitem[Monteith \emph{et~al.}(2016)Monteith, Brunner, Djagaeva, Bielecki,
  Deutsch, and Saxton]{Monteith2016}
C.~E. Monteith, M.~E. Brunner, I.~Djagaeva, A.~M. Bielecki, J.~M. Deutsch and
  W.~M. Saxton, \emph{Biophysical Journal}, 2016, \textbf{110},
  2053--2065\relax
\mciteBstWouldAddEndPuncttrue
\mciteSetBstMidEndSepPunct{\mcitedefaultmidpunct}
{\mcitedefaultendpunct}{\mcitedefaultseppunct}\relax
\EndOfBibitem
\bibitem[Stein \emph{et~al.}(2021)Stein, De~Canio, Lauga, Shelley, and
  Goldstein]{PhysRevLett.126.028103}
D.~B. Stein, G.~De~Canio, E.~Lauga, M.~J. Shelley and R.~E. Goldstein,
  \emph{Phys. Rev. Lett.}, 2021, \textbf{126}, 028103\relax
\mciteBstWouldAddEndPuncttrue
\mciteSetBstMidEndSepPunct{\mcitedefaultmidpunct}
{\mcitedefaultendpunct}{\mcitedefaultseppunct}\relax
\EndOfBibitem
\bibitem[Lin \emph{et~al.}(2010)Lin, Shenoy, Hu, and Bai]{LIN20101043}
Y.~Lin, V.~Shenoy, B.~Hu and L.~Bai, \emph{Biophysical Journal}, 2010,
  \textbf{99}, 1043--1052\relax
\mciteBstWouldAddEndPuncttrue
\mciteSetBstMidEndSepPunct{\mcitedefaultmidpunct}
{\mcitedefaultendpunct}{\mcitedefaultseppunct}\relax
\EndOfBibitem
\bibitem[Howard(2001)]{Howard2001}
J.~Howard, \emph{Mechanics of Motor Proteins and the Cytoskeleton}, Sinauer
  Associates, Sunderland, Massachusetts, 2001\relax
\mciteBstWouldAddEndPuncttrue
\mciteSetBstMidEndSepPunct{\mcitedefaultmidpunct}
{\mcitedefaultendpunct}{\mcitedefaultseppunct}\relax
\EndOfBibitem
\bibitem[Phillips \emph{et~al.}(2012)Phillips, Kondev, Theriot, and
  Garcia]{phillips2012physical}
R.~Phillips, J.~Kondev, J.~Theriot and H.~G. Garcia, \emph{Physical Biology of
  the Cell}, Garland Science, 2nd edn., 2012\relax
\mciteBstWouldAddEndPuncttrue
\mciteSetBstMidEndSepPunct{\mcitedefaultmidpunct}
{\mcitedefaultendpunct}{\mcitedefaultseppunct}\relax
\EndOfBibitem
\bibitem[Kumar \emph{et~al.}(2025)Kumar, Inamdar, Pullarkat, and
  Menon]{kumar2025forcesscalecell}
K.~V. Kumar, M.~M. Inamdar, P.~A. Pullarkat and G.~I. Menon, \emph{Forces at
  the scale of the cell}, 2025, \url{https://arxiv.org/abs/2512.08311}\relax
\mciteBstWouldAddEndPuncttrue
\mciteSetBstMidEndSepPunct{\mcitedefaultmidpunct}
{\mcitedefaultendpunct}{\mcitedefaultseppunct}\relax
\EndOfBibitem
\bibitem[Joanny and Prost(2009)]{joanny2009active}
J.-F. Joanny and J.~Prost, \emph{HFSP journal}, 2009, \textbf{3}, 94--104\relax
\mciteBstWouldAddEndPuncttrue
\mciteSetBstMidEndSepPunct{\mcitedefaultmidpunct}
{\mcitedefaultendpunct}{\mcitedefaultseppunct}\relax
\EndOfBibitem
\bibitem[Diotallevi(2007)]{ffffa456249a4bed89485e12d11cd89c}
F.~Diotallevi, \emph{external PhD, WU}, Wageningen University, 2007\relax
\mciteBstWouldAddEndPuncttrue
\mciteSetBstMidEndSepPunct{\mcitedefaultmidpunct}
{\mcitedefaultendpunct}{\mcitedefaultseppunct}\relax
\EndOfBibitem
\bibitem[Cannon(2000)]{Cannon2000}
R.~E. Cannon, in \emph{Acetobacter xylinum --- Biotechnology and Food
  Technology}, ed. N.~Eynard and J.~Teissi{\'e}, Springer Berlin Heidelberg,
  Berlin, Heidelberg, 2000, pp. 104--107\relax
\mciteBstWouldAddEndPuncttrue
\mciteSetBstMidEndSepPunct{\mcitedefaultmidpunct}
{\mcitedefaultendpunct}{\mcitedefaultseppunct}\relax
\EndOfBibitem
\bibitem[Lopes \emph{et~al.}(2025)Lopes, Winterstrain, Caballero, Chardac,
  Alvarado, Cusi, Dalal, Kelly, Stehnach, Goode, Fai, Hagan, Norton, and
  Duclos]{Lopes2025EmergenceOA}
J.~D. Lopes, B.~Winterstrain, F.~Caballero, A.~Chardac, I.~Alvarado, A.~T.~D.
  Cusi, S.~A. Dalal, G.~Kelly, M.~R. Stehnach, B.~L. Goode, T.~G. Fai, M.~F.
  Hagan, M.~M. Norton and G.~Duclos, author, 2025\relax
\mciteBstWouldAddEndPuncttrue
\mciteSetBstMidEndSepPunct{\mcitedefaultmidpunct}
{\mcitedefaultendpunct}{\mcitedefaultseppunct}\relax
\EndOfBibitem
\bibitem[Bunea and Taboryski(2020)]{mi11121048}
A.-I. Bunea and R.~Taboryski, \emph{Micromachines}, 2020, \textbf{11},
  year\relax
\mciteBstWouldAddEndPuncttrue
\mciteSetBstMidEndSepPunct{\mcitedefaultmidpunct}
{\mcitedefaultendpunct}{\mcitedefaultseppunct}\relax
\EndOfBibitem
\bibitem[Ganguly and Raj(2026)]{GANGULY2026204529}
S.~Ganguly and K.~Raj, \emph{European Journal of Mechanics - B/Fluids}, 2026,
  \textbf{119}, 204529\relax
\mciteBstWouldAddEndPuncttrue
\mciteSetBstMidEndSepPunct{\mcitedefaultmidpunct}
{\mcitedefaultendpunct}{\mcitedefaultseppunct}\relax
\EndOfBibitem
\bibitem[Farhadifar \emph{et~al.}(2020)Farhadifar, Yu, Fabig, Wu, Stein,
  Rockman, Müller-Reichert, Shelley, and Needleman]{10.7554/eLife.55877}
R.~Farhadifar, C.-H. Yu, G.~Fabig, H.-Y. Wu, D.~B. Stein, M.~Rockman,
  T.~Müller-Reichert, M.~J. Shelley and D.~J. Needleman, \emph{eLife}, 2020,
  \textbf{9}, e55877\relax
\mciteBstWouldAddEndPuncttrue
\mciteSetBstMidEndSepPunct{\mcitedefaultmidpunct}
{\mcitedefaultendpunct}{\mcitedefaultseppunct}\relax
\EndOfBibitem
\bibitem[Chakrabarti \emph{et~al.}(2024)Chakrabarti, Rachh, Shvartsman, and
  Shelley]{doi:10.1073/pnas.2405114121}
B.~Chakrabarti, M.~Rachh, S.~Y. Shvartsman and M.~J. Shelley, \emph{Proceedings
  of the National Academy of Sciences}, 2024, \textbf{121}, e2405114121\relax
\mciteBstWouldAddEndPuncttrue
\mciteSetBstMidEndSepPunct{\mcitedefaultmidpunct}
{\mcitedefaultendpunct}{\mcitedefaultseppunct}\relax
\EndOfBibitem
\bibitem[Batchelor(1970)]{Batchelor_1970}
G.~K. Batchelor, \emph{Journal of Fluid Mechanics}, 1970, \textbf{44},
  419–440\relax
\mciteBstWouldAddEndPuncttrue
\mciteSetBstMidEndSepPunct{\mcitedefaultmidpunct}
{\mcitedefaultendpunct}{\mcitedefaultseppunct}\relax
\EndOfBibitem
\bibitem[Jeffery(1922)]{jeffery1922motion}
G.~B. Jeffery, \emph{Proceedings of the Royal Society of London. Series A,
  Containing Papers of a Mathematical and Physical Character}, 1922,
  \textbf{102}, 161--179\relax
\mciteBstWouldAddEndPuncttrue
\mciteSetBstMidEndSepPunct{\mcitedefaultmidpunct}
{\mcitedefaultendpunct}{\mcitedefaultseppunct}\relax
\EndOfBibitem
\bibitem[Dutta \emph{et~al.}(2024)Dutta, Farhadifar, Lu, Kabacaoğlu,
  Blackwell, Stein, Lakonishok, Gelfand, Shvartsman, and Shelley]{Dutta2024}
S.~Dutta, R.~Farhadifar, W.~Lu, G.~Kabacaoğlu, R.~Blackwell, D.~B. Stein,
  M.~Lakonishok, V.~I. Gelfand, S.~Y. Shvartsman and M.~J. Shelley,
  \emph{Nature Physics}, 2024, \textbf{20}, 666--674\relax
\mciteBstWouldAddEndPuncttrue
\mciteSetBstMidEndSepPunct{\mcitedefaultmidpunct}
{\mcitedefaultendpunct}{\mcitedefaultseppunct}\relax
\EndOfBibitem
\bibitem[Jain \emph{et~al.}(2025)Jain, Chakrabarti, Farhadifar, Gavis, Shelley,
  and Shvartsman]{PRXLife.3.023007}
O.~Jain, B.~Chakrabarti, R.~Farhadifar, E.~R. Gavis, M.~J. Shelley and S.~Y.
  Shvartsman, \emph{PRX Life}, 2025, \textbf{3}, 023007\relax
\mciteBstWouldAddEndPuncttrue
\mciteSetBstMidEndSepPunct{\mcitedefaultmidpunct}
{\mcitedefaultendpunct}{\mcitedefaultseppunct}\relax
\EndOfBibitem
\bibitem[Keller \emph{et~al.}(1975)Keller, Wu, and Brennen]{keller1975traction}
S.~Keller, T.~Wu and C.~Brennen, 1975\relax
\mciteBstWouldAddEndPuncttrue
\mciteSetBstMidEndSepPunct{\mcitedefaultmidpunct}
{\mcitedefaultendpunct}{\mcitedefaultseppunct}\relax
\EndOfBibitem
\bibitem[Ishikawa \emph{et~al.}(2020)Ishikawa, Pedley, Drescher, and
  Goldstein]{ishikawa2020stability}
T.~Ishikawa, T.~Pedley, K.~Drescher and R.~E. Goldstein, \emph{Journal of Fluid
  Mechanics}, 2020, \textbf{903}, A11\relax
\mciteBstWouldAddEndPuncttrue
\mciteSetBstMidEndSepPunct{\mcitedefaultmidpunct}
{\mcitedefaultendpunct}{\mcitedefaultseppunct}\relax
\EndOfBibitem
\bibitem[Kanale \emph{et~al.}(2022)Kanale, Ling, Guo, F{\"u}rthauer, and
  Kanso]{kanale2022spontaneous}
A.~V. Kanale, F.~Ling, H.~Guo, S.~F{\"u}rthauer and E.~Kanso, \emph{Proceedings
  of the National Academy of Sciences}, 2022, \textbf{119}, e2214413119\relax
\mciteBstWouldAddEndPuncttrue
\mciteSetBstMidEndSepPunct{\mcitedefaultmidpunct}
{\mcitedefaultendpunct}{\mcitedefaultseppunct}\relax
\EndOfBibitem
\bibitem[Weady \emph{et~al.}(2022)Weady, Shelley, and Stein]{WEADY2022110937}
S.~Weady, M.~J. Shelley and D.~B. Stein, \emph{Journal of Computational
  Physics}, 2022, \textbf{457}, 110937\relax
\mciteBstWouldAddEndPuncttrue
\mciteSetBstMidEndSepPunct{\mcitedefaultmidpunct}
{\mcitedefaultendpunct}{\mcitedefaultseppunct}\relax
\EndOfBibitem
\bibitem[Kim and Karrila(2005)]{kim2005microhydrodynamics}
S.~Kim and S.~Karrila, \emph{Microhydrodynamics: Principles and Selected
  Applications}, Dover Publications, 2005\relax
\mciteBstWouldAddEndPuncttrue
\mciteSetBstMidEndSepPunct{\mcitedefaultmidpunct}
{\mcitedefaultendpunct}{\mcitedefaultseppunct}\relax
\EndOfBibitem
\bibitem[Rachh \emph{et~al.}(2017)Rachh, Kl{\"o}ckner, and
  O'Neil]{rachh2017fast}
M.~Rachh, A.~Kl{\"o}ckner and M.~O'Neil, \emph{Journal of Computational
  Physics}, 2017, \textbf{345}, 706--731\relax
\mciteBstWouldAddEndPuncttrue
\mciteSetBstMidEndSepPunct{\mcitedefaultmidpunct}
{\mcitedefaultendpunct}{\mcitedefaultseppunct}\relax
\EndOfBibitem
\bibitem[Corona \emph{et~al.}(2017)Corona, Greengard, Rachh, and
  Veerapaneni]{corona2017integral}
E.~Corona, L.~Greengard, M.~Rachh and S.~Veerapaneni, \emph{Journal of
  Computational Physics}, 2017, \textbf{332}, 504--519\relax
\mciteBstWouldAddEndPuncttrue
\mciteSetBstMidEndSepPunct{\mcitedefaultmidpunct}
{\mcitedefaultendpunct}{\mcitedefaultseppunct}\relax
\EndOfBibitem
\bibitem[Glotzer \emph{et~al.}(1997)Glotzer, Saffrich, Glotzer, and
  Ephrussi]{glotzer1997cytoplasmic}
J.~B. Glotzer, R.~Saffrich, M.~Glotzer and A.~Ephrussi, \emph{Current Biology},
  1997, \textbf{7}, 326--337\relax
\mciteBstWouldAddEndPuncttrue
\mciteSetBstMidEndSepPunct{\mcitedefaultmidpunct}
{\mcitedefaultendpunct}{\mcitedefaultseppunct}\relax
\EndOfBibitem
\bibitem[Quinlan(2016)]{quinlan2016cytoplasmic}
M.~E. Quinlan, \emph{Annual review of cell and developmental biology}, 2016,
  \textbf{32}, 173--195\relax
\mciteBstWouldAddEndPuncttrue
\mciteSetBstMidEndSepPunct{\mcitedefaultmidpunct}
{\mcitedefaultendpunct}{\mcitedefaultseppunct}\relax
\EndOfBibitem
\bibitem[Ganguly \emph{et~al.}(2012)Ganguly, Williams, Palacios, and
  Goldstein]{ganguly2012cytoplasmic}
S.~Ganguly, L.~S. Williams, I.~M. Palacios and R.~E. Goldstein,
  \emph{Proceedings of the National Academy of Sciences}, 2012, \textbf{109},
  15109--15114\relax
\mciteBstWouldAddEndPuncttrue
\mciteSetBstMidEndSepPunct{\mcitedefaultmidpunct}
{\mcitedefaultendpunct}{\mcitedefaultseppunct}\relax
\EndOfBibitem
\bibitem[Khuc~Trong \emph{et~al.}(2015)Khuc~Trong, Doerflinger, Dunkel,
  St~Johnston, and Goldstein]{khuc2015cortical}
P.~Khuc~Trong, H.~Doerflinger, J.~Dunkel, D.~St~Johnston and R.~E. Goldstein,
  \emph{Elife}, 2015, \textbf{4}, e06088\relax
\mciteBstWouldAddEndPuncttrue
\mciteSetBstMidEndSepPunct{\mcitedefaultmidpunct}
{\mcitedefaultendpunct}{\mcitedefaultseppunct}\relax
\EndOfBibitem
\bibitem[Lauga and Powers(2009)]{Lauga_2009}
E.~Lauga and T.~R. Powers, \emph{Reports on Progress in Physics}, 2009,
  \textbf{72}, 096601\relax
\mciteBstWouldAddEndPuncttrue
\mciteSetBstMidEndSepPunct{\mcitedefaultmidpunct}
{\mcitedefaultendpunct}{\mcitedefaultseppunct}\relax
\EndOfBibitem
\bibitem[Li \emph{et~al.}(1989)Li, Wang, and Knox]{Li1989}
Y.~Li, F.~H. Wang and R.~B. Knox, \emph{Protoplasma}, 1989, \textbf{149},
  57--63\relax
\mciteBstWouldAddEndPuncttrue
\mciteSetBstMidEndSepPunct{\mcitedefaultmidpunct}
{\mcitedefaultendpunct}{\mcitedefaultseppunct}\relax
\EndOfBibitem
\bibitem[Das \emph{et~al.}(2026)Das, Zhu, Bonnet, and
  Veerapaneni]{das2026squirmersarbitraryshapeslip}
K.~Das, H.~Zhu, M.~Bonnet and S.~Veerapaneni, \emph{Squirmers with arbitrary
  shape and slip: modeling, simulation, and optimization}, 2026,
  \url{https://arxiv.org/abs/2602.19336}\relax
\mciteBstWouldAddEndPuncttrue
\mciteSetBstMidEndSepPunct{\mcitedefaultmidpunct}
{\mcitedefaultendpunct}{\mcitedefaultseppunct}\relax
\EndOfBibitem
\bibitem[Stone and Samuel(1996)]{PhysRevLett.77.4102}
H.~A. Stone and A.~D.~T. Samuel, \emph{Phys. Rev. Lett.}, 1996, \textbf{77},
  4102--4104\relax
\mciteBstWouldAddEndPuncttrue
\mciteSetBstMidEndSepPunct{\mcitedefaultmidpunct}
{\mcitedefaultendpunct}{\mcitedefaultseppunct}\relax
\EndOfBibitem
\bibitem[Kimura \emph{et~al.}(2017)Kimura, Mamane, Sasaki, Sato, Takagi,
  Niwayama, Hufnagel, Shimamoto, Joanny,
  Uchida,\emph{et~al.}]{kimura2017endoplasmic}
K.~Kimura, A.~Mamane, T.~Sasaki, K.~Sato, J.~Takagi, R.~Niwayama, L.~Hufnagel,
  Y.~Shimamoto, J.-F. Joanny, S.~Uchida \emph{et~al.}, \emph{Nature Cell
  Biology}, 2017, \textbf{19}, 399--406\relax
\mciteBstWouldAddEndPuncttrue
\mciteSetBstMidEndSepPunct{\mcitedefaultmidpunct}
{\mcitedefaultendpunct}{\mcitedefaultseppunct}\relax
\EndOfBibitem
\bibitem[Sulerud \emph{et~al.}(2020)Sulerud, Sami, Li, Kloxin, Oakey, and
  Gatlin]{sulerud2020microtubule}
T.~Sulerud, A.~B. Sami, G.~Li, A.~Kloxin, J.~Oakey and J.~Gatlin,
  \emph{Molecular biology of the cell}, 2020, \textbf{31}, 2791--2802\relax
\mciteBstWouldAddEndPuncttrue
\mciteSetBstMidEndSepPunct{\mcitedefaultmidpunct}
{\mcitedefaultendpunct}{\mcitedefaultseppunct}\relax
\EndOfBibitem
\bibitem[Gong \emph{et~al.}(2021)Gong, Mathijssen, Bryant, and
  Prakash]{gong2021engineering}
X.~Gong, A.~J. Mathijssen, Z.~Bryant and M.~Prakash, \emph{Physical Review
  Fluids}, 2021, \textbf{6}, 123104\relax
\mciteBstWouldAddEndPuncttrue
\mciteSetBstMidEndSepPunct{\mcitedefaultmidpunct}
{\mcitedefaultendpunct}{\mcitedefaultseppunct}\relax
\EndOfBibitem
\bibitem[Lopes \emph{et~al.}(2025)Lopes, Winterstrain, Caballero, Chardac,
  Alvarado, Cusi, Dalal, Kelly, Stehnach, Goode, Fai, Hagan, Norton, and
  Duclos]{lopes2025emergenceantichemotacticflockingactive}
J.~D. Lopes, B.~Winterstrain, F.~Caballero, A.~Chardac, I.~Alvarado, A.~T.
  Cusi, S.~Dalal, G.~Kelly, M.~R. Stehnach, B.~L. Goode, T.~G. Fai, M.~F.
  Hagan, M.~M. Norton and G.~Duclos, \emph{Emergence of Anti-chemotactic
  Flocking in Active Biomimetic Colloids}, 2025,
  \url{https://arxiv.org/abs/2505.17394}\relax
\mciteBstWouldAddEndPuncttrue
\mciteSetBstMidEndSepPunct{\mcitedefaultmidpunct}
{\mcitedefaultendpunct}{\mcitedefaultseppunct}\relax
\EndOfBibitem
\end{mcitethebibliography}

%
\end{document}